\gdef\isLaptop{F}
\gdef\PSfonts{T}
\magnification\magstep1

\newdimen\papwidth
\newdimen\papheight
\newskip\beforesectionskipamount  
\newskip\sectionskipamount 
\def\sectionskip{\vskip\sectionskipamount}
\def\beforesectionskip{\vskip\beforesectionskipamount}
\papwidth=16truecm
\if F\isLaptop
\papheight=22truecm
\voffset=0.4truecm
\hoffset=0.4truecm
\else
\papheight=16truecm
\voffset=-1.5truecm
\hoffset=0.4truecm
\fi
\hsize=\papwidth
\vsize=\papheight
\nopagenumbers
\headline={\ifnum\pageno>1 {\hss\tenrm-\ \folio\ -\hss} \else
{\hfill}\fi}
\newdimen\texpscorrection
\texpscorrection=0.15truecm 

\def\sectionsize{\twelvepoint}
\def\sectiontype{\bf}
\def\subsectionsize{}
\def\subsectiontype{\bf}
\def\em{\sl}
\newfam\truecmsy
\newfam\truecmr
\newfam\msbfam
\newfam\scriptfam
\newfam\truecmsy
\newskip\ttglue 
\if T\isLaptop
\papheight=11.5truecm
\fi
\if F\PSfonts
\font\twelverm=cmr12
\font\tenrm=cmr10
\font\eightrm=cmr8
\font\sevenrm=cmr7
\font\sixrm=cmr6
\font\fiverm=cmr5

\font\twelvebf=cmbx12
\font\tenbf=cmbx10
\font\eightbf=cmbx8
\font\sevenbf=cmbx7
\font\sixbf=cmbx6
\font\fivebf=cmbx5

\font\twelveit=cmti12
\font\tenit=cmti10
\font\eightit=cmti8
\font\sevenit=cmti7
\font\sixit=cmti6
\font\fiveit=cmti5

\font\twelvesl=cmsl12
\font\tensl=cmsl10
\font\eightsl=cmsl8
\font\sevensl=cmsl7
\font\sixsl=cmsl6
\font\fivesl=cmsl5

\font\twelvei=cmmi12
\font\teni=cmmi10
\font\eighti=cmmi8
\font\seveni=cmmi7
\font\sixi=cmmi6
\font\fivei=cmmi5

\font\twelvesy=cmsy10	at	12pt
\font\tensy=cmsy10
\font\eightsy=cmsy8
\font\sevensy=cmsy7
\font\sixsy=cmsy6
\font\fivesy=cmsy5
\font\twelvetruecmsy=cmsy10	at	12pt
\font\tentruecmsy=cmsy10
\font\eighttruecmsy=cmsy8
\font\seventruecmsy=cmsy7
\font\sixtruecmsy=cmsy6
\font\fivetruecmsy=cmsy5

\font\twelvetruecmr=cmr12
\font\tentruecmr=cmr10
\font\eighttruecmr=cmr8
\font\seventruecmr=cmr7
\font\sixtruecmr=cmr6
\font\fivetruecmr=cmr5

\font\twelvebf=cmbx12
\font\tenbf=cmbx10
\font\eightbf=cmbx8
\font\sevenbf=cmbx7
\font\sixbf=cmbx6
\font\fivebf=cmbx5

\font\twelvett=cmtt12
\font\tentt=cmtt10
\font\eighttt=cmtt8

\font\twelveex=cmex10	at	12pt
\font\tenex=cmex10

\font\twelvemsb=msbm10	at	12pt
\font\tenmsb=msbm10
\font\eightmsb=msbm8
\font\sevenmsb=msbm7
\font\sixmsb=msbm6
\font\fivemsb=msbm5

\font\twelvescr=eusm10 at 12pt
\font\tenscr=eusm10
\font\eightscr=eusm8
\font\sevenscr=eusm7
\font\sixscr=eusm6
\font\fivescr=eusm5
\fi
\if T\PSfonts
\font\twelverm=ptmr	at	12pt
\font\tenrm=ptmr	at	10pt
\font\eightrm=ptmr	at	8pt
\font\sevenrm=ptmr	at	7pt
\font\sixrm=ptmr	at	6pt
\font\fiverm=ptmr	at	5pt

\font\twelvebf=ptmb	at	12pt
\font\tenbf=ptmb	at	10pt
\font\eightbf=ptmb	at	8pt
\font\sevenbf=ptmb	at	7pt
\font\sixbf=ptmb	at	6pt
\font\fivebf=ptmb	at	5pt

\font\twelveit=ptmri	at	12pt
\font\tenit=ptmri	at	10pt
\font\eightit=ptmri	at	8pt
\font\sevenit=ptmri	at	7pt
\font\sixit=ptmri	at	6pt
\font\fiveit=ptmri	at	5pt

\font\twelvesl=ptmro	at	12pt
\font\tensl=ptmro	at	10pt
\font\eightsl=ptmro	at	8pt
\font\sevensl=ptmro	at	7pt
\font\sixsl=ptmro	at	6pt
\font\fivesl=ptmro	at	5pt

\font\twelvei=cmmi12
\font\teni=cmmi10
\font\eighti=cmmi8
\font\seveni=cmmi7
\font\sixi=cmmi6
\font\fivei=cmmi5

\font\twelvesy=cmsy10	at	12pt
\font\tensy=cmsy10
\font\eightsy=cmsy8
\font\sevensy=cmsy7
\font\sixsy=cmsy6
\font\fivesy=cmsy5
\font\twelvetruecmsy=cmsy10	at	12pt
\font\tentruecmsy=cmsy10
\font\eighttruecmsy=cmsy8
\font\seventruecmsy=cmsy7
\font\sixtruecmsy=cmsy6
\font\fivetruecmsy=cmsy5

\font\twelvetruecmr=cmr12
\font\tentruecmr=cmr10
\font\eighttruecmr=cmr8
\font\seventruecmr=cmr7
\font\sixtruecmr=cmr6
\font\fivetruecmr=cmr5

\font\twelvebf=cmbx12
\font\tenbf=cmbx10
\font\eightbf=cmbx8
\font\sevenbf=cmbx7
\font\sixbf=cmbx6
\font\fivebf=cmbx5

\font\twelvett=cmtt12
\font\tentt=cmtt10
\font\eighttt=cmtt8

\font\twelveex=cmex10	at	12pt
\font\tenex=cmex10

\font\twelvemsb=msbm10	at	12pt
\font\tenmsb=msbm10
\font\eightmsb=msbm8
\font\sevenmsb=msbm7
\font\sixmsb=msbm6
\font\fivemsb=msbm5

\font\twelvescr=eusm10 at 12pt
\font\tenscr=eusm10
\font\eightscr=eusm8
\font\sevenscr=eusm7
\font\sixscr=eusm6
\font\fivescr=eusm5
\fi
\def\eightpoint{\def\rm{\fam0\eightrm}%
\textfont0=\eightrm
  \scriptfont0=\sixrm
  \scriptscriptfont0=\fiverm 
\textfont1=\eighti
  \scriptfont1=\sixi
  \scriptscriptfont1=\fivei 
\textfont2=\eightsy
  \scriptfont2=\sixsy
  \scriptscriptfont2=\fivesy 
\textfont3=\tenex
  \scriptfont3=\tenex
  \scriptscriptfont3=\tenex 
\textfont\itfam=\eightit
  \scriptfont\itfam=\sixit
  \scriptscriptfont\itfam=\fiveit 
  \def\it{\fam\itfam\eightit}%
\textfont\slfam=\eightsl
  \scriptfont\slfam=\sixsl
  \scriptscriptfont\slfam=\fivesl 
  \def\sl{\fam\slfam\eightsl}%
\textfont\ttfam=\eighttt
  \def\tt{\fam\ttfam\eighttt}%
\textfont\bffam=\eightbf
  \scriptfont\bffam=\sixbf
  \scriptscriptfont\bffam=\fivebf
  \def\bf{\fam\bffam\eightbf}%
\textfont\scriptfam=\eightscr
  \scriptfont\scriptfam=\sixscr
  \scriptscriptfont\scriptfam=\fivescr
  \def\script{\fam\scriptfam\eightscr}%
\textfont\msbfam=\eightmsb
  \scriptfont\msbfam=\sixmsb
  \scriptscriptfont\msbfam=\fivemsb
  \def\bb{\fam\msbfam\eightmsb}%
\textfont\truecmr=\eighttruecmr
  \scriptfont\truecmr=\sixtruecmr
  \scriptscriptfont\truecmr=\fivetruecmr
  \def\truerm{\fam\truecmr\eighttruecmr}%
\textfont\truecmsy=\eighttruecmsy
  \scriptfont\truecmsy=\sixtruecmsy
  \scriptscriptfont\truecmsy=\fivetruecmsy
\tt \ttglue=.5em plus.25em minus.15em 
\normalbaselineskip=9pt
\setbox\strutbox=\hbox{\vrule height7pt depth2pt width0pt}%
\normalbaselines
\rm
}

\def\tenpoint{\def\rm{\fam0\tenrm}%
\textfont0=\tenrm
  \scriptfont0=\sevenrm
  \scriptscriptfont0=\fiverm 
\textfont1=\teni
  \scriptfont1=\seveni
  \scriptscriptfont1=\fivei 
\textfont2=\tensy
  \scriptfont2=\sevensy
  \scriptscriptfont2=\fivesy 
\textfont3=\tenex
  \scriptfont3=\tenex
  \scriptscriptfont3=\tenex 
\textfont\itfam=\tenit
  \scriptfont\itfam=\sevenit
  \scriptscriptfont\itfam=\fiveit 
  \def\it{\fam\itfam\tenit}%
\textfont\slfam=\tensl
  \scriptfont\slfam=\sevensl
  \scriptscriptfont\slfam=\fivesl 
  \def\sl{\fam\slfam\tensl}%
\textfont\ttfam=\tentt
  \def\tt{\fam\ttfam\tentt}%
\textfont\bffam=\tenbf
  \scriptfont\bffam=\sevenbf
  \scriptscriptfont\bffam=\fivebf
  \def\bf{\fam\bffam\tenbf}%
\textfont\scriptfam=\tenscr
  \scriptfont\scriptfam=\sevenscr
  \scriptscriptfont\scriptfam=\fivescr
  \def\script{\fam\scriptfam\tenscr}%
\textfont\msbfam=\tenmsb
  \scriptfont\msbfam=\sevenmsb
  \scriptscriptfont\msbfam=\fivemsb
  \def\bb{\fam\msbfam\tenmsb}%
\textfont\truecmr=\tentruecmr
  \scriptfont\truecmr=\seventruecmr
  \scriptscriptfont\truecmr=\fivetruecmr
  \def\truerm{\fam\truecmr\tentruecmr}%
\textfont\truecmsy=\tentruecmsy
  \scriptfont\truecmsy=\seventruecmsy
  \scriptscriptfont\truecmsy=\fivetruecmsy
\tt \ttglue=.5em plus.25em minus.15em 
\normalbaselineskip=12pt
\setbox\strutbox=\hbox{\vrule height8.5pt depth3.5pt width0pt}%
\normalbaselines
\rm
}

\def\twelvepoint{\def\rm{\fam0\twelverm}%
\textfont0=\twelverm
  \scriptfont0=\tenrm
  \scriptscriptfont0=\eightrm 
\textfont1=\twelvei
  \scriptfont1=\teni
  \scriptscriptfont1=\eighti 
\textfont2=\twelvesy
  \scriptfont2=\tensy
  \scriptscriptfont2=\eightsy 
\textfont3=\twelveex
  \scriptfont3=\twelveex
  \scriptscriptfont3=\twelveex 
\textfont\itfam=\twelveit
  \scriptfont\itfam=\tenit
  \scriptscriptfont\itfam=\eightit 
  \def\it{\fam\itfam\twelveit}%
\textfont\slfam=\twelvesl
  \scriptfont\slfam=\tensl
  \scriptscriptfont\slfam=\eightsl 
  \def\sl{\fam\slfam\twelvesl}%
\textfont\ttfam=\twelvett
  \def\tt{\fam\ttfam\twelvett}%
\textfont\bffam=\twelvebf
  \scriptfont\bffam=\tenbf
  \scriptscriptfont\bffam=\eightbf
  \def\bf{\fam\bffam\twelvebf}%
\textfont\scriptfam=\twelvescr
  \scriptfont\scriptfam=\tenscr
  \scriptscriptfont\scriptfam=\eightscr
  \def\script{\fam\scriptfam\twelvescr}%
\textfont\msbfam=\twelvemsb
  \scriptfont\msbfam=\tenmsb
  \scriptscriptfont\msbfam=\eightmsb
  \def\bb{\fam\msbfam\twelvemsb}%
\textfont\truecmr=\twelvetruecmr
  \scriptfont\truecmr=\tentruecmr
  \scriptscriptfont\truecmr=\eighttruecmr
  \def\truerm{\fam\truecmr\twelvetruecmr}%
\textfont\truecmsy=\twelvetruecmsy
  \scriptfont\truecmsy=\tentruecmsy
  \scriptscriptfont\truecmsy=\eighttruecmsy
\tt \ttglue=.5em plus.25em minus.15em 
\setbox\strutbox=\hbox{\vrule height7pt depth2pt width0pt}%
\normalbaselineskip=15pt
\normalbaselines
\rm
}
%
\fontdimen16\tensy=2.7pt
\fontdimen13\tensy=4.3pt
\fontdimen17\tensy=2.7pt
\fontdimen14\tensy=4.3pt
\fontdimen18\tensy=4.3pt
\fontdimen16\eightsy=2.7pt
\fontdimen13\eightsy=4.3pt
\fontdimen17\eightsy=2.7pt
\fontdimen14\eightsy=4.3pt
\fontdimen18\eightsy=4.3pt
%
\def\hexnumber#1{\ifcase#1 0\or1\or2\or3\or4\or5\or6\or7\or8\or9\or
 A\or B\or C\or D\or E\or F\fi}
\mathcode`\=="3\hexnumber\truecmr3D
\mathchardef\not="3\hexnumber\truecmsy36
\mathcode`\+="2\hexnumber\truecmr2B
\mathcode`\(="4\hexnumber\truecmr28
\mathcode`\)="5\hexnumber\truecmr29
\mathcode`\!="5\hexnumber\truecmr21
\mathcode`\(="4\hexnumber\truecmr28
\mathcode`\)="5\hexnumber\truecmr29

\def\dot{\mathaccent"0\hexnumber\truecmr5F }
\def\Phi{\mathchar"0\hexnumber\truecmr08 }
\def\Gamma {\mathchar"0\hexnumber\truecmr00 }
\def\Delta {\mathchar"0\hexnumber\truecmr01 }
\def\Theta {\mathchar"0\hexnumber\truecmr02 }
\def\Lambda{\mathchar"0\hexnumber\truecmr03 }
\def\Xi {\mathchar"0\hexnumber\truecmr04 }
\def\Pi{\mathchar"0\hexnumber\truecmr05 }
\def\Sigma{\mathchar"0\hexnumber\truecmr06 }
\def\Upsilon {\mathchar"0\hexnumber\truecmr07 }
\def\Phi {\mathchar"0\hexnumber\truecmr08 }
\def\Psi {\mathchar"0\hexnumber\truecmr09 }
\def\Omega{\mathchar"0\hexnumber\truecmr0A }
\newcount\EQNcount \EQNcount=1
\newcount\CLAIMcount \CLAIMcount=1
\newcount\SECTIONcount \SECTIONcount=0
\newcount\SUBSECTIONcount \SUBSECTIONcount=1
\def\ifff(#1,#2,#3){\ifundefined{#1#2}%
\expandafter\xdef\csname #1#2\endcsname{#3}\else%
\immediate\write16{!!!!!doubly defined #1,#2}\fi}
\def\NEWDEF #1,#2,#3 {\ifff({#1},{#2},{#3})}
\def\actualnumber{\number\SECTIONcount}
\def\EQ(#1){\lmargin(#1)\eqno\tag(#1)}
\def\NR(#1){&\lmargin(#1)\tag(#1)\cr}  
\def\tag(#1){\lmargin(#1)({\rm \actualnumber}.\number\EQNcount)
 \NEWDEF e,#1,(\actualnumber.\number\EQNcount)
\global\advance\EQNcount by 1
}
\def\SECT(#1)#2\par{\lmargin(#1)\SECTION#2\par
\NEWDEF s,#1,{\actualnumber}
}
\def\SUBSECT(#1)#2\par{\lmargin(#1)
\SUBSECTION#2\par 
\NEWDEF s,#1,{\actualnumber.\number\SUBSECTIONcount}
}
\def\CLAIM #1(#2) #3\par{
\vskip.1in\medbreak\noindent
{\lmargin(#2)\bf #1\ \actualnumber.\number\CLAIMcount.} {\sl #3}\par
\NEWDEF c,#2,{#1\ \actualnumber.\number\CLAIMcount}
\global\advance\CLAIMcount by 1
\ifdim\lastskip<\medskipamount
\removelastskip\penalty55\medskip\fi}
\def\CLAIMNONR #1(#2) #3\par{
\vskip.1in\medbreak\noindent
{\lmargin(#2)\bf #1.} {\sl #3}\par
\NEWDEF c,#2,{#1}
\global\advance\CLAIMcount by 1
\ifdim\lastskip<\medskipamount
\removelastskip\penalty55\medskip\fi}
\def\SECTION#1\par{\vskip0pt plus.3\vsize\penalty-75
    \vskip0pt plus -.3\vsize
    \global\advance\SECTIONcount by 1
    \beforesectionskip\noindent
{\sectionsize\sectiontype \actualnumber.\ #1}
    \EQNcount=1
    \CLAIMcount=1
    \SUBSECTIONcount=1
    \nobreak\sectionskip\noindent}
\def\SECTIONNONR#1\par{\vskip0pt plus.3\vsize\penalty-75
    \vskip0pt plus -.3\vsize
    \global\advance\SECTIONcount by 1
    \beforesectionskip\noindent
{\sectionsize\sectiontype  #1}
     \EQNcount=1
     \CLAIMcount=1
     \SUBSECTIONcount=1
     \nobreak\sectionskip\noindent}
\def\SUBSECTION#1\par{\vskip0pt plus.2\vsize\penalty-75%
    \vskip0pt plus -.2\vsize%
    \beforesectionskip\noindent%
{\subsectionsize\subsectiontype \actualnumber.\number\SUBSECTIONcount.\ #1}
    \global\advance\SUBSECTIONcount by 1
    \nobreak\sectionskip\noindent}
\def\SUBSECTIONNONR#1\par{\vskip0pt plus.2\vsize\penalty-75
    \vskip0pt plus -.2\vsize
\beforesectionskip\noindent
{\subsectionsize\subsectiontype #1}
    \nobreak\sectionskip\noindent\noindent}
\def\ifundefined#1{\expandafter\ifx\csname#1\endcsname\relax}
\def\equ(#1){\ifundefined{e#1}$\spadesuit$#1\else\csname e#1\endcsname\fi}
\def\clm(#1){\ifundefined{c#1}$\spadesuit$#1\else\csname c#1\endcsname\fi}
\def\sec(#1){\ifundefined{s#1}$\spadesuit$#1
\else Section \csname s#1\endcsname\fi}
\let\endarg=\par
\def\finish{\def\endarg{\par\endgroup}}
\def\start{\endarg\begingroup}

 \def\beginFROM{\start\parskip=0pt\vskip\baselineskip
\def\finish{\def\endarg{\egroup\par\endgroup}}
  \vbox\bgroup\obeylines\eightpoint\em\finish}

\def\ABSTRACT#1\par{
\vskip 1in {\noindent\sectionsize\sectiontype Abstract.} #1 \par}

\def\TODAY{\number\day~\ifcase\month\or January \or February \or March \or
April \or May \or June
\or July \or August \or September \or October \or November \or December \fi
\number\year\timecount=\number\time
\divide\timecount by 60
}
\newcount\timecount
\def\DRAFT{\def\lmargin(##1){\strut\vadjust{\kern-\strutdepth
\vtop to \strutdepth{
\baselineskip\strutdepth\vss\rlap{\kern-1.2 truecm\eightpoint{##1}}}}}
\font\footfont=cmti7
\footline={{\footfont \hfil File:\jobname, \TODAY,  \number\timecount h}}
}
\newbox\strutboxJPE
\setbox\strutboxJPE=\hbox{\strut}
\def\subitem#1#2\par{\vskip\baselineskip\vskip-\ht\strutboxJPE{\item{#1}#2}}
\gdef\strutdepth{\dp\strutbox}
\def\lmargin(#1){}
\def\period{\unskip.\spacefactor3000 { }}
%
%
\newbox\noboxJPE
\newbox\byboxJPE
\newbox\paperboxJPE
\newbox\yrboxJPE
\newbox\jourboxJPE
\newbox\pagesboxJPE
\newbox\volboxJPE
\newbox\preprintboxJPE
\newbox\toappearboxJPE
\newbox\bookboxJPE
\newbox\bybookboxJPE
\newbox\publisherboxJPE
\newbox\inprintboxJPE
\def\refclearJPE{
   \setbox\noboxJPE=\null             \gdef\isnoJPE{F}
   \setbox\byboxJPE=\null             \gdef\isbyJPE{F}
   \setbox\paperboxJPE=\null          \gdef\ispaperJPE{F}
   \setbox\yrboxJPE=\null             \gdef\isyrJPE{F}
   \setbox\jourboxJPE=\null           \gdef\isjourJPE{F}
   \setbox\pagesboxJPE=\null          \gdef\ispagesJPE{F}
   \setbox\volboxJPE=\null            \gdef\isvolJPE{F}
   \setbox\preprintboxJPE=\null       \gdef\ispreprintJPE{F}
   \setbox\toappearboxJPE=\null       \gdef\istoappearJPE{F}
   \setbox\inprintboxJPE=\null        \gdef\isinprintJPE{F}
   \setbox\bookboxJPE=\null           \gdef\isbookJPE{F}  \gdef\isinbookJPE{F}
     
   \setbox\bybookboxJPE=\null         \gdef\isbybookJPE{F}
   \setbox\publisherboxJPE=\null      \gdef\ispublisherJPE{F}
     
}
\def\widestlabel#1{\setbox0=\hbox{#1\enspace}\refindent=\wd0\relax}
\def\ref{\refclearJPE\bgroup}
\def\no   {\egroup\gdef\isnoJPE{T}\setbox\noboxJPE=\hbox\bgroup}
\def\by   {\egroup\gdef\isbyJPE{T}\setbox\byboxJPE=\hbox\bgroup}
\def\paper{\egroup\gdef\ispaperJPE{T}\setbox\paperboxJPE=\hbox\bgroup}
\def\yr{\egroup\gdef\isyrJPE{T}\setbox\yrboxJPE=\hbox\bgroup}
\def\jour{\egroup\gdef\isjourJPE{T}\setbox\jourboxJPE=\hbox\bgroup}
\def\pages{\egroup\gdef\ispagesJPE{T}\setbox\pagesboxJPE=\hbox\bgroup}
\def\vol{\egroup\gdef\isvolJPE{T}\setbox\volboxJPE=\hbox\bgroup\bf}
\def\preprint{\egroup\gdef
\ispreprintJPE{T}\setbox\preprintboxJPE=\hbox\bgroup}
\def\toappear{\egroup\gdef
\istoappearJPE{T}\setbox\toappearboxJPE=\hbox\bgroup}
\def\inprint{\egroup\gdef
\isinprintJPE{T}\setbox\inprintboxJPE=\hbox\bgroup}
\def\book{\egroup\gdef\isbookJPE{T}\setbox\bookboxJPE=\hbox\bgroup\em}
\def\publisher{\egroup\gdef
\ispublisherJPE{T}\setbox\publisherboxJPE=\hbox\bgroup}
\def\inbook{\egroup\gdef\isinbookJPE{T}\setbox\bookboxJPE=\hbox\bgroup\em}
\def\bybook{\egroup\gdef\isbybookJPE{T}\setbox\bybookboxJPE=\hbox\bgroup}
\newdimen\refindent
\refindent=5em
\def\endref{\egroup \sfcode`.=1000
 \if T\isnoJPE
 \hangindent\refindent\hangafter=1
      \noindent\hbox to\refindent{[\unhbox\noboxJPE\unskip]\hss}\ignorespaces
     \else  \noindent    \fi
 \if T\isbyJPE    \unhbox\byboxJPE\unskip: \fi
 \if T\ispaperJPE \unhbox\paperboxJPE\unskip\period \fi
 \if T\isbookJPE {\it\unhbox\bookboxJPE\unskip}\if T\ispublisherJPE, \else.
\fi\fi
 \if T\isinbookJPE In {\it\unhbox\bookboxJPE\unskip}\if T\isbybookJPE,
\else\period \fi\fi
 \if T\isbybookJPE  (\unhbox\bybookboxJPE\unskip)\period \fi
 \if T\ispublisherJPE \unhbox\publisherboxJPE\unskip \if T\isjourJPE, \else\if
T\isyrJPE \  \else\period \fi\fi\fi
 \if T\istoappearJPE (To appear)\period \fi
 \if T\ispreprintJPE Pre\-print\period \fi
 \if T\isjourJPE    \unhbox\jourboxJPE\unskip\ \fi
 \if T\isvolJPE     \unhbox\volboxJPE\unskip\if T\ispagesJPE, \else\ \fi\fi
 \if T\ispagesJPE   \unhbox\pagesboxJPE\unskip\  \fi
 \if T\isyrJPE      (\unhbox\yrboxJPE\unskip)\period \fi
 \if T\isinprintJPE (in print)\period \fi
\filbreak
}
\def\hexnumber#1{\ifcase#1 0\or1\or2\or3\or4\or5\or6\or7\or8\or9\or
 A\or B\or C\or D\or E\or F\fi}
\textfont\msbfam=\tenmsb
\scriptfont\msbfam=\sevenmsb
\scriptscriptfont\msbfam=\fivemsb
\mathchardef\varkappa="0\hexnumber\msbfam7B
\newcount\FIGUREcount \FIGUREcount=0
\newdimen\figcenter
\def\fig(#1){\ifundefined{fig#1}%
\global\advance\FIGUREcount by 1%
\NEWDEF fig,#1,{Fig.\ \number\FIGUREcount}
\immediate\write16{ FIG \number\FIGUREcount : #1}
\fi
\csname fig#1\endcsname\relax}
\def\figure #1 #2 #3 #4\cr{\null%
\ifundefined{fig#1}%
\global\advance\FIGUREcount by 1%
\NEWDEF fig,#1,{Fig.\ \number\FIGUREcount}
\immediate\write16{  FIG \number\FIGUREcount : #1}
\fi
{\goodbreak\figcenter=\hsize\relax
\advance\figcenter by -#3truecm
\divide\figcenter by 2
\midinsert\vskip #2truecm\noindent\hskip\figcenter
\includegraphics{#1}\vskip 0.8truecm\noindent \vbox{\eightpoint\noindent
{\bf\fig(#1)}: #4}\endinsert}}
\def\figurewithtex #1 #2 #3 #4 #5\cr{\null%
\ifundefined{fig#1}%
\global\advance\FIGUREcount by 1%
\NEWDEF fig,#1,{Fig.\ \number\FIGUREcount}
\immediate\write16{ FIG \number\FIGUREcount: #1}
\fi
{\goodbreak\figcenter=\hsize\relax
\advance\figcenter by -#4truecm
\divide\figcenter by 2
\midinsert\vskip #3truecm\noindent\hskip\figcenter
\includegraphics{#1}{\hskip\texpscorrection\input #2 }\vskip 0.8truecm\noindent \vbox{\eightpoint\noindent
{\bf\fig(#1)}: #5}\endinsert}}
\def\figurewithtexplus #1 #2 #3 #4 #5 #6\cr{\null%
\ifundefined{fig#1}%
\global\advance\FIGUREcount by 1%
\NEWDEF fig,#1,{Fig.\ \number\FIGUREcount}
\immediate\write16{ FIG \number\FIGUREcount: #1}
\fi
{\goodbreak\figcenter=\hsize\relax
\advance\figcenter by -#4truecm
\divide\figcenter by 2
\midinsert\vskip #3truecm\noindent\hskip\figcenter
\includegraphics{#1}{\hskip\texpscorrection\input #2 }\vskip #5truecm\noindent \vbox{\eightpoint\noindent
{\bf\fig(#1)}: #6}\endinsert}}
\catcode`@=11
\def\footnote#1{\let\@sf\empty 
  \ifhmode\edef\@sf{\spacefactor\the\spacefactor}\/\fi
  #1\@sf\vfootnote{#1}}
\def\vfootnote#1{\insert\footins\bgroup\eightpoint
  \interlinepenalty\interfootnotelinepenalty
  \splittopskip\ht\strutbox 
  \splitmaxdepth\dp\strutbox \floatingpenalty\@MM
  \leftskip\z@skip \rightskip\z@skip \spaceskip\z@skip \xspaceskip\z@skip
  \textindent{#1}\footstrut\futurelet\next\fo@t}
\def\fo@t{\ifcat\bgroup\noexpand\next \let\next\f@@t
  \else\let\next\f@t\fi \next}
\def\f@@t{\bgroup\aftergroup\@foot\let\next}
\def\f@t#1{#1\@foot}
\def\@foot{\strut\egroup}
\def\footstrut{\vbox to\splittopskip{}}
\skip\footins=\bigskipamount 
\count\footins=1000 
\dimen\footins=8in 
\catcode`@=12 

\def\BB{{\script B}}
\def\CC{{\script C}}

\def\HH{{\script H}}
\def\LL{{\script L}}
\def\MM{{\script M}}
\def\NN{{\script N}}
\def\OO{{\script O}}

\def\HALF{{\textstyle{1\over 2}}}

\def\QED{\hfill\smallskip
         \line{$\hfill{\vcenter{\vbox{\hrule height 0.2pt
	\hbox{\vrule width 0.2pt height 1.8ex \kern 1.8ex
		\vrule width 0.2pt}
	\hrule height 0.2pt}}}$
               \ \ \ \ \ \ }
         \bigskip}
\def\real{{\bf R}}

\def\Re{{\rm Re\,}}

\def\PROOF{\medskip\noindent{\bf Proof.\ }}
\def\REMARK{\medskip\noindent{\bf Remark.\ }}
\def\LIKEREMARK#1{\medskip\noindent{\bf #1.\ }}
\tenpoint
\normalbaselineskip=5.25mm
\baselineskip=5.25mm
\parskip=10pt
\beforesectionskipamount=24pt plus8pt minus8pt
\sectionskipamount=3pt plus1pt minus1pt
\def\em{\it}
\normalbaselineskip=12pt
\baselineskip=12pt
\parskip=0pt
\parindent=22.222pt
\beforesectionskipamount=24pt plus0pt minus6pt
\sectionskipamount=7pt plus3pt minus0pt
\overfullrule=0pt
\hfuzz=2pt
\nopagenumbers
\headline={\ifnum\pageno>1 {\hss\tenrm-\ \folio\ -\hss} \else
{\hfill}\fi}
\if F\PSfonts
\font\titlefont=cmbx10 scaled\magstep2

\font\toplinefont=cmr10
\font\pagenumberfont=cmr10
\let\tenpoint=\rm
\else
\font\titlefont=ptmb at 14 pt

\font\toplinefont=cmcsc10
\font\pagenumberfont=ptmb at 10pt
\fi
\newdimen\itemindent\itemindent=1.5em

\def\textindent#1{\indent\llap{#1\enspace}\ignorespaces}
\def\item{\par\noindent
\hangindent\itemindent\hangafter=1\relax
\setitemmark}
\def\setitemindent#1{\setbox0=\hbox{\ignorespaces#1\unskip\enspace}%
\itemindent=\wd0\relax
\message{|\string\setitemindent: Mark width modified to hold
         |`\string#1' plus an \string\enspace\space gap. }%
}
\def\setitemmark#1{\checkitemmark{#1}%
\hbox to\itemindent{\hss#1\enspace}\ignorespaces}
\def\checkitemmark#1{\setbox0=\hbox{\enspace#1}%
\ifdim\wd0>\itemindent
   \message{|\string\item: Your mark `\string#1' is too wide. }%
\fi}
\def\SECTION#1\par{\vskip0pt plus.2\vsize\penalty-75
    \vskip0pt plus -.2\vsize
    \global\advance\SECTIONcount by 1
    \beforesectionskip\noindent
{\sectionsize\sectiontype \actualnumber.\ #1}
    \EQNcount=1
    \CLAIMcount=1
    \SUBSECTIONcount=1
    \nobreak\sectionskip\noindent}
\def\Equ(#1){Eq.\equ(#1) }
\def\frac#1#2{{#1\over #2}}
\def\e{e}
\def\hspace#1{\quad}
\def\THEOREM(#1){\CLAIM Theorem(#1)}
\def\LEMMA(#1){\CLAIM Lemma(#1)}
\def\section#1{\SECTION #1}
\def\subsection#1{\SUBSECTION #1}
\def\label#1{}
\def\Lemma(#1){\clm(#1)}
\def\Proposition(#1){\clm(#1)}
\def\Theorem(#1){\clm(#1)}
\def\DEFINITION(#1){\CLAIM Definition(#1)}
\def\PROPOSITION(#1){\CLAIM Proposition(#1)}
\def\COROLLARY(#1){\CLAIM Corollary(#1)}
\def\CITE[#1]{[#1]}
\let\mathcal=\cal
\def\Equ(#1){Eq.\equ(#1)}
\def\ITEM[#1]{\ref\no #1}
\def\REF#1 #2 #3 #4 #5 #6\par{\by #1
\paper #2
\jour #3
\vol #4
\pages #5
\yr #6\endref}
\def\BOOK#1 #2 #3 #4 #5\par{\by #1
\book #2
\publisher #3, #4
\yr #5\endref}
\def\INBOOK#1 #2 #3 #4 #5 #6\par{\by #1
\paper #2
\inbook #3 \bybook #4 \publisher #5
\yr #6\endref}
\def\TOAPPEAR#1 #2 #3\par{\by #1\paper #2\jour #3\toappear \endref}
\def\TODO#1 #2\par{\by #1 \paper #2 (In preparation)\endref}
\def\PREPRINT#1 #2 #3\par{\by #1\paper #2 (Preprint) \yr #3\endref} 
\def\CMP{{Commun. Math. Phys.}}
\def\JSP{{J. Stat. Phys.}}
\def\JMP{{J. Math. Phys.}}

\def\LMP{{Lett. Math. Phys.}}
\def\JP{{Jak\v si\'c,~V. and C.-A.~Pillet}}
\let\truett=\tt
\fontdimen3\tentt=2pt\fontdimen4\tentt=2pt
\def\tt{\hfill\break\null\kern -2truecm\truett **** }
\def\BB{{\cal B}}
\def\CC{{\cal C}}
\def\DD{{\cal D}}
\def\MM{{\cal M}}
\def\FF{{\cal F}}
\def\SS{{\cal S}}
\def\WW{{\cal W}}
\def\pa{\partial}
\def\supp{{\rm supp}}
\def\const{{\rm const.}}
\def\lff{\bigl( \| \Lop f \| + \| f \| \bigr )}
\def\nn{{n+1}}
\def\HH{{\mathcal H}}
\def\NN{{\cal N}}
\def\L{{\rm L}}
\def\R{{\rm R}}

\let\para=\S
\def\S{{\rm S}}

\def\as{{\rm as}}
\def\d{{\rm d}}
\def\dpp{{\rm dp}}
\def\dq{{\rm dq}}
\def\dr{{\rm dr}}
\def\dx{{\rm dx}}
\def\dy{{\rm dy}}
\def\dt{{\rm d}t}
\def\ds{{\rm d}s}
\def\FF{{\cal F}}

\def\YY{{\cal Y}}
\def\iii{\int_0^\infty \!\!\d z\, }
\let\epsilon=\varepsilon
\let\kappa=\varkappa
\let\theta=\vartheta
\def\hidekappa#1{}
\def\norm{\vert\kern-0.1em\vert\kern-0.1em\vert}
\def\bnorm{\big\vert\kern-0.1em\big\vert\kern-0.1em\big\vert}
\def\smm{\sum_{m=1}^M}
\def\Lm{{\L,m}}
\def\Rm{{\R,m}}
\def\im{{i,m}}
\def\T{{\rm T}}
\def\mlist{m=1,\dots,M}
\def\D{{\cal M}}
\def\zl{W_\Lm}
\def\zr{W_\Rm}
\def\zi{W_{i,m}}
\def\wlim{\mathop{\rm \hbox{w--lim}}}
\def\smu{_\HH}
\def\Lop{K}
\def\Calkop{{\cal K}}
\def\quatre{4}
\def\TTH0{T^t_{\HH_0}}
\def\TTHH{T^t_{\HH}}


\setitemindent{\bf H3)}
\def\ifff(#1,#2,#3){\ifundefined{#1#2}\expandafter\xdef\csname #1#2\endcsname{#3}\write16{defining #1#2}\fi}
\expandafter\xdef\csname
ehamil\endcsname{(2.1)}
\expandafter\xdef\csname
escalar\endcsname{(2.2)}
\expandafter\xdef\csname
ecoupham\endcsname{(2.3)}
\expandafter\xdef\csname
eeqmo1\endcsname{(2.4)}
\expandafter\xdef\csname
eeqmo2\endcsname{(2.5)}
\expandafter\xdef\csname
ecov1\endcsname{(2.6)}
\expandafter\xdef\csname
efluc-diss\endcsname{(2.7)}
\expandafter\xdef\csname
eexp\endcsname{(2.8)}
\expandafter\xdef\csname
eeqmo3\endcsname{(2.9)}
\expandafter\xdef\csname
c1\endcsname{Theorem\ 2.1}
\expandafter\xdef\csname
c2\endcsname{Theorem\ 2.2}
\expandafter\xdef\csname
elimit1\endcsname{(2.10)}
\expandafter\xdef\csname
cc1\endcsname{Corollary\ 2.3}
\expandafter\xdef\csname
esde\endcsname{(3.1)}
\expandafter\xdef\csname
ediv\endcsname{(3.2)}
\expandafter\xdef\csname
ethed\endcsname{(3.3)}
\expandafter\xdef\csname
esol\endcsname{(3.4)}
\expandafter\xdef\csname
egena\endcsname{(3.5)}
\expandafter\xdef\csname
eTT\endcsname{(3.6)}
\expandafter\xdef\csname
eeffective\endcsname{(3.7)}
\expandafter\xdef\csname
ebeta\endcsname{(3.8)}
\expandafter\xdef\csname
ehh0\endcsname{(3.9)}
\expandafter\xdef\csname
ealpha\endcsname{(3.10)}
\expandafter\xdef\csname
egen1\endcsname{(3.11)}
\expandafter\xdef\csname
ezetadef\endcsname{(3.12)}
\expandafter\xdef\csname
eLiouv\endcsname{(3.13)}
\expandafter\xdef\csname
eposi\endcsname{(3.14)}
\expandafter\xdef\csname
eone\endcsname{(3.15)}
\expandafter\xdef\csname
cI.1\endcsname{Lemma\ 3.1}
\expandafter\xdef\csname
eKol\endcsname{(3.16)}
\expandafter\xdef\csname
csmef\endcsname{Corollary\ 3.2}
\expandafter\xdef\csname
ctrprob\endcsname{Corollary\ 3.3}
\expandafter\xdef\csname
cI.com\endcsname{Proposition\ 3.4}
\expandafter\xdef\csname
cschw\endcsname{Proposition\ 3.5}
\expandafter\xdef\csname
cI.posnew\endcsname{Proposition\ 3.6}
\expandafter\xdef\csname
eim\endcsname{(3.17)}
\expandafter\xdef\csname
eouff\endcsname{(3.18)}
\expandafter\xdef\csname
econditions\endcsname{(3.19)}
\expandafter\xdef\csname
csimplicity\endcsname{Lemma\ 3.7}
\expandafter\xdef\csname
eresol\endcsname{(3.20)}
\expandafter\xdef\csname
cuniqueness\endcsname{Theorem\ 3.8}
\expandafter\xdef\csname
ebirerg\endcsname{(3.21)}
\expandafter\xdef\csname
etttt\endcsname{(3.22)}
\expandafter\xdef\csname
ecesaro\endcsname{(3.23)}
\expandafter\xdef\csname
cmixing\endcsname{Proposition\ 3.9}
\expandafter\xdef\csname
ewlim\endcsname{(3.24)}
\expandafter\xdef\csname
edecc\endcsname{(3.25)}
\expandafter\xdef\csname
el1t\endcsname{(3.26)}
\expandafter\xdef\csname
estarseq\endcsname{(3.27)}
\expandafter\xdef\csname
er0\endcsname{(3.28)}
\expandafter\xdef\csname
erind1\endcsname{(3.29)}
\expandafter\xdef\csname
ewell\endcsname{(3.30)}
\expandafter\xdef\csname
egen1repeat\endcsname{(4.1)}
\expandafter\xdef\csname
ezetadefrep\endcsname{(4.2)}
\expandafter\xdef\csname
eLiouvrep\endcsname{(4.3)}
\expandafter\xdef\csname
epositive\endcsname{(4.4)}
\expandafter\xdef\csname
ealldefs\endcsname{(4.5)}
\expandafter\xdef\csname
eh1\endcsname{(4.6)}
\expandafter\xdef\csname
chor1\endcsname{Theorem\ 4.1}
\expandafter\xdef\csname
epineq\endcsname{(4.7)}
\expandafter\xdef\csname
epineq2\endcsname{(4.8)}
\expandafter\xdef\csname
eqineq\endcsname{(4.9)}
\expandafter\xdef\csname
eqineq2\endcsname{(4.10)}
\expandafter\xdef\csname
erineq\endcsname{(4.11)}
\expandafter\xdef\csname
chor2\endcsname{Proposition\ 4.2}
\expandafter\xdef\csname
estartinduct\endcsname{(4.12)}
\expandafter\xdef\csname
ecomm1\endcsname{(4.13)}
\expandafter\xdef\csname
ecomm2\endcsname{(4.14)}
\expandafter\xdef\csname
einductp\endcsname{(4.15)}
\expandafter\xdef\csname
einductq\endcsname{(4.16)}
\expandafter\xdef\csname
cpseudo\endcsname{Proposition\ 4.3}
\expandafter\xdef\csname
cq\endcsname{Lemma\ 4.4}
\expandafter\xdef\csname
eh5\endcsname{(4.17)}
\expandafter\xdef\csname
epainful\endcsname{(4.18)}
\expandafter\xdef\csname
eh6\endcsname{(4.19)}
\expandafter\xdef\csname
ealmostlast\endcsname{(4.20)}
\expandafter\xdef\csname
ccompact\endcsname{Proposition\ 4.5}
\expandafter\xdef\csname
ecall\endcsname{(4.21)}
\expandafter\xdef\csname
csmooth\endcsname{Proposition\ 4.6}
\expandafter\xdef\csname
ccinfty\endcsname{Proposition\ 4.7}
\expandafter\xdef\csname
ccore\endcsname{Theorem\ 4.8}
\expandafter\xdef\csname
egym\endcsname{(4.22)}
\expandafter\xdef\csname
efog\endcsname{(4.23)}
\expandafter\xdef\csname
ehorm\endcsname{(4.24)}
\expandafter\xdef\csname
enep\endcsname{(4.25)}
\expandafter\xdef\csname
cl2\endcsname{Lemma\ A.1}
\expandafter\xdef\csname
ebdiv\endcsname{(A.1)}
\expandafter\xdef\csname
cl2w\endcsname{Lemma\ A.2}
\expandafter\xdef\csname
elist\endcsname{(B.1)}
\expandafter\xdef\csname
elamclam\endcsname{(B.2)}
\expandafter\xdef\csname
clamclam\endcsname{Lemma\ B.1}
\expandafter\xdef\csname
ebs1\endcsname{(B.3)}
\expandafter\xdef\csname
ebs2\endcsname{(B.4)}
\expandafter\xdef\csname
cbs\endcsname{Proposition\ B.2}
\expandafter\xdef\csname
eb3\endcsname{(B.5)}
\expandafter\xdef\csname
eb4\endcsname{(B.6)}
\expandafter\xdef\csname
cintbound\endcsname{Lemma\ B.3}
\expandafter\xdef\csname
einsint\endcsname{(B.7)}
\expandafter\xdef\csname
eex1\endcsname{(B.8)}
\expandafter\xdef\csname
ecalpha\endcsname{(B.9)}
\expandafter\xdef\csname
eclon\endcsname{(C.1)}
\def\NEWDEF #1,#2,#3 {\ifff({#1},{#2},{#3})}
%
%

{\titlefont{\centerline{Non-Equilibrium Statistical Mechanics of 
Anharmonic Chains  }}}
\vskip 0.5truecm
{\titlefont{\centerline{Coupled to Two Heat Baths at Different Temperatures }}}
\vskip 0.5truecm
{\it{\centerline{J.-P. Eckmann${}^{1,2}$, C.-A. Pillet${}^{3,4}$ and
L. Rey-Bellet${}^{2}$ }}
\vskip 0.3truecm
{\eightpoint
\centerline{${}^1$D\'ept.~de Physique Th\'eorique, Universit\'e de Gen\`eve,
CH-1211 Gen\`eve 4, Switzerland}
\centerline{${}^2$Section de Math\'ematiques, Universit\'e de Gen\`eve,
CH-1211 Gen\`eve 4, Switzerland}
\centerline{${}^3$PHYMAT, Universit\'e de Toulon,
F-83957 La Garde Cedex, France}
\centerline{${}^4$CPT-CNRS Luminy, F-13288 Marseille Cedex 09}
}}
\vskip 0.5truecm\headline
{\ifnum\pageno>1 {\toplinefont Anharmonic Chains
Coupled to Two Heat Baths}
\hfill{\pagenumberfont\folio}\fi}
\vskip 3cm
{\eightpoint\narrower\baselineskip 11pt
\LIKEREMARK{Abstract} We study the statistical mechanics of a
finite-dimensional non-linear Hamiltonian system (a chain of
{\it anharmonic} oscillators) coupled to two heat baths (described by wave
equations).  Assuming that the initial conditions of the heat baths
are distributed
according to the Gibbs measures at two {\it different}
temperatures 
we study the dynamics of the oscillators. Under suitable assumptions
on the potential and on the coupling between the chain and the heat
baths, we prove the existence of an invariant measure for {\it any}
temperature difference, {\it i.e.}, we prove the existence of {\it steady
states}. Furthermore, if the temperature difference is
sufficiently small, we prove that the invariant measure is {\it
unique} and {\it mixing}.  In particular, we develop new techniques
for proving the existence of 
invariant measures for random processes on a non-compact phase
space. These techniques are based on an extension of  the commutator
method of H\"ormander used in the study of hypoelliptic differential 
operators.
}

\vfill\eject

%
\SECTION Introduction

In this paper, we consider the non-equilibrium statistical mechanics
of a finite-dimensional non-linear Hamiltonian system coupled to two
infinite heat baths which are at {\em different} temperatures. We show
that under certain conditions on the initial data the system goes to a
unique non-equilibrium {\em steady state}.

To put this new result into perspective, we situate it among other
results in equilibrium and non-equilibrium statistical
mechanics.
First of all, for the case of only {\em one} heat bath
one expects of course ``return to equilibrium.''
This problem has a long history, and a proof of return to equilibrium 
under quite general conditions on the {\em non-linear} small system
and its coupling to the heat bath has been recently obtained in
[JP1-4].%
Viewed from context of our present problem, the main simplifying
feature of the one-bath problem is that the final state can be
guessed, {\it a priori}, to be the familiar Boltzmann distribution.

For the case of two heat baths, there are no results of such
generality available, among other things precisely because one cannot
guess in general what the steady state is going to be. Since we are
dealing with systems on a non-compact phase space and without energy
conservation, there is nothing like an SRB Ansatz for our problem [GC].
Worse, even the existence of any stationary state is not obvious
at all. The only notable exceptions are problems where the small system and its
coupling to the heat baths are linear. Then the problem can be formulated in
terms of Gaussian measures, and approach to a steady state has been
proved in this case in [RLL], [CL] and [OL] for Markovian heat baths
and in [SL] for the general case.

Our approach in the present paper will consist in using the spirit of
[FKM] and [FK] to give a microscopic derivation of the equations of
motion: under suitable assumptions, we will reduce the study of the
dynamics of the coupled system (an infinite dimensional Hamiltonian
system) to the study of a random finite dimensional dynamical system.
However, we will not achieve the generality of [JP1-4]. 
Each heat bath is an infinite dimensional linear Hamiltonian system,
in our case it will be chosen as the
classical field theory associated with the wave 
equation. The small system is a non-linear Hamiltonian system with 
an arbitrary (but finite) number of
degrees of freedom, in our case it is chosen
as a chain of anharmonic oscillators with nearest neighbor couplings.
The potential must be of quadratic type near infinite energies.
The two heat baths are coupled respectively to the first and the last
particle of the chain. The initial conditions of the heat baths will
be distributed according to thermal equilibrium at inverse
temperatures $\beta _\L$, $\beta _\R$. Integrating the variables of
the heat baths leads to a system of random integro-differential
equations: the generalized Langevin equations. They differ from the 
Newton equations of motion by the addition of two kinds of force,
on one hand there is a (random) force exerted by the heat baths on
the chain of oscillators and on the other hand there is a dissipative
force with memory which describe the genuine retro-action from the heat bath
on the small system. We will choose the couplings between the baths and the
chain such that the random forces exerted by the baths have an
exponentially decaying covariance. With this assumption (see [Tr]),
the resulting equations are quasi-Markovian. By this, we mean that one
can introduce a finite number of auxiliary variables in such a way 
that the evolution of the chain, together with these variables, is
described by a system of Markovian stochastic differential equations.

With this set-up, we are led to a classical problem in probability
theory: the study of invariant measures for diffusion processes.
For our problem, the main difficulties are as follows: the phase space is
not compact and the resulting diffusion process is degenerate and not
self-adjoint.\footnote{${}^1$}{The diffusion is non-selfadjoint
because the diffusion process is not time-reversal invariant. But
the generator $L$ satisfies $L^*T=TL$ where $T$ changes the sign of
the momenta.} 
The standard
techniques used to prove the existence of invariant measures do not
seem to work in our case and, in this paper, we develop new methods
to solve this problem, which rely on methods of spectral analysis.
Our proof of existence is based on a compactness
argument, as often in the proof of existence of invariant measures. 
More precisely we will prove that the generator of the diffusion process,
a second order differential operator, given in our problem, has compact
resolvent, in a suitably chosen Hilbert space. This is done by
generalizing the commutator method of H\"ormander, [H\"o], used in the
study of hypoelliptic operators. Similar methods have been used 
to study the spectrum of Schr\"odinger operators with magnetic fields,
see [HM], [He], and [HN].

The restriction to a chain is mostly for convenience. Other geometries
can be accommodated with our methods, and the number of heat baths is not
restricted to two. Furthermore, the techniques developed in this paper
can be applied to other interesting models of non-equilibrium
statistical mechanics, for example, an electric field
acting on a system of particles [R-B, in preparation].

%
\SECTION Description of the Model and Derivation of the Effective Equations

In this section we define a model of two heat baths coupled
to a small system, and derive the stochastic equations which describe
the time evolution of the small system. The heat baths are classical
field theories associated with the wave equation, the small system is a
chain of oscillators and the coupling between them is linear in the
field.

We begin the description of the model by defining the ``small''
system.
It is a chain of $d$-dimensional anharmonic
oscillators. The phase space
of the chain is $\real^{2dn}$ with $n$ and $d$ arbitrary and its
dynamics is described by a 
$\CC^{\infty}$ Hamiltonian function of the form
$$
H_\S (p,q)\,=\,\sum_{j=1}^n \frac{p_j^2}{2} + \sum_{j=1}^n U_j(q_j)
+ \sum_{i=1}^{n-1} U^{(2)}_{i}(q_i,q_{i+1})\,\equiv\,\sum_{j=1}^n \frac{p_j^2}{2} +V(q)~,
\EQ(hamil)
$$
where $q=(q_1,\dots,q_n)$, $p=(p_1,\dots,p_n)$,
with $p_i$, $q_i \in \real^{d}$.

The potential energy will be assumed ``quadratic $+$ bounded'' in the
following sense.
We let $\FF$ denote the space of $\CC^\infty $ functions on $\real^{dn}$ such
that for all multi-indices $\alpha $ and all $F\in\FF$, the quantity
$
\partial^\alpha F(q)
$ is bounded  uniformly in $q\in\real^{dn}$.
Then our hypotheses are
\medskip
\item{\bf H1)}Behavior at infinity: We assume that $V$ is of the form
$$
V(q)\,=\,\HALF\bigl (q-a,{\bf Q}(q-a)\bigr )+F(q)~,
$$
where ${\bf Q}$ is a positive definite ($dn\times dn$) matrix, $a$ is a
vector, and $\partial_{q_i^{(\nu)}}F\in\FF$ for $i=1,\dots,n$ and
$\nu=1,\dots,d$.
\item{\bf H2)}Coupling: Each of the ($d\times d$) matrices
$$
\D_{i,i+1}(q)\,\equiv\,\nabla_{q_i}\nabla_{q_{i+1}}U^{(2)}_i(q_i,q_{i+1})~,\quad i=1,\dots,n-1~,
$$
is
either uniformly positive or negative definite.
\REMARK The first hypothesis makes sure the particles do not ``fly
away.'' The second hypothesis makes sure that the nearest neighbor interaction
can transmit energy. As such, this
condition is of the hypoelliptic type. 
\LIKEREMARK{Example}A typical case (in dimension $d$)
covered by these hypotheses is given by
$$
U_j(q)\,=\, q^2 +5 \sin\bigl (\sqrt{1+q^2}\bigr )~,\quad
U^{(2)}_i(q,q')\,=\, (q-q')^2 +\sin\bigl (\sqrt{1+(q-q')^2}\bigr )/(2d)~.
$$

As a model of a heat bath we consider the classical field theory associated
with the $d$-dimensional wave equation.
The field $\varphi$ and its conjugate momentum field $\pi$ are elements
of the real Hilbert space $\HH={\rm H}^1_{\real}(\real^{d}) \oplus
\L ^2_{\real}(\real^{d})$ which is the completion of $\CC_0^\infty
(\real^d)\oplus\CC_0^\infty
(\real^d)$
with respect to the norm defined by the scalar product:
$$
\left(\left(\matrix{\varphi\cr\pi\cr}\right),
\left(\matrix{\varphi\cr\pi\cr}\right)\right)_\HH
\,=\,\int {\dx}\,\left(|\nabla\varphi(x)|^2+|\pi(x)|^2\right)~.
\EQ(scalar)
$$
The Hamilton function of the free heat bath is
$$
H_{\rm B}(\varphi,\pi)=\frac{1}{2}\int {\rm
dx}\,\left(|\nabla\varphi(x)|^2+|\pi(x)|^2\right)  ~,
$$
and the corresponding equation of motion is the ordinary wave equation
which we write in the form
$$
\left(\matrix{{\dot\varphi}(t)\cr{\dot\pi}(t)\cr}\right)\,=\,
\LL \left(\matrix{\varphi\cr\pi\cr}\right)~,
$$
where
$$
\LL\,\equiv\,\left(\matrix{0&1\cr\Delta&0\cr}\right)~.
$$

Let us turn to the coupling between the chain and the heat baths.
The baths will be called ``$\L$'' and ``$\R$'', the left bath
couples to the coordinate $q_1$ and the right bath couples to the
other end of the chain ($q_n$).
Since we consider two heat baths,
the phase space of the coupled system, for finite energy
configurations, is $\real^{2dn} \times {\cal{H}} \times {\cal{H}}$
and
its Hamiltonian will be chosen as
$$\eqalign{
H(p,q,\varphi_{\L},\pi_{\L},\varphi_{\R},\pi_{\R})\,&=\, H_\S (p,q) +
H_{\rm B}(\varphi_\L,\pi_\L) + H_{\rm B}(\varphi_\R,\pi_\R) \cr
&+ 
\hidekappa{\kappa_\L } q_1 \cdot \int {\dx}\, \rho_\L (x) \nabla \varphi_\L(x)
+ \hidekappa{\kappa_\R}  q_n \cdot \int {\dx} \,\rho_\R (x) \nabla
\varphi_\R(x)~. 
}\EQ(coupham)
$$
Here, the \hidekappa{$\kappa_i$, $i\in\{\L,\R\}$ 
are real coupling constants and the }$\rho_j(x)\in\L^1(\real^d)$ are
charge densities which we assume for simplicity to be
spherically symmetric functions.
The choice of the Hamiltonian
\Equ(coupham) is motivated by the dipole approximation of classical
electrodynamics. For notational purposes we use in the sequel the shorthand
$$
\phi_i\equiv \left(\matrix{\varphi_i\cr\pi_i\cr}\right)~.
$$
We set
$\alpha_i=\left(\alpha_{i}^{(1)},\dots,\alpha_{i}^{(d)}\right)$, 
$i\in\{\L,\R\}$, with
$$
{\widehat \alpha}_{i}^{(\nu)}(k)\equiv \left(\matrix{{-ik^{(\nu)}}
{\widehat \rho}_i(k)/k^2\cr 0\cr}\right)~.
$$
Here and in the sequel the ``hat'' means the Fourier transform
$$
{\widehat f}(k)\,\equiv\,\frac{1}{(2\pi)^{d/2}}\int {\dx}\,f(x)\e^{-ik\cdot
x}~.
$$
With this notation the Hamiltonian becomes
$$
H(p,q,\phi_\L,\phi_\R)\,=\,
H_\S (p,q)+H_{\rm B}(\phi_\L)+H_{\rm B}(\phi_\R)+q_1 \cdot
(\phi_\L,\alpha _\L)_{\HH}+q_n \cdot
(\phi_\R,\alpha _\R)_{\HH}~,
$$
where $H_{{\rm B}}(\phi)=\HALF\|\phi\|_{\HH}^2$.
We next study the equations of motions. They take the form
$$\eqalign{
{\dot q}_j (t) \,&=\, p_j (t)~, \qquad j =1,\dots,n~, \cr
{\dot p}_1(t) \,&=\, -\nabla_{q_1}V(q(t))-\hidekappa{\kappa_\L}
(\phi_\L(t),\alpha_\L )_{\HH}~, \cr 
{\dot p}_j (t) \,&=\, -\nabla_{q_j }V(q(t))~, \qquad  j =2,\dots,n-1
{}~,\cr
{\dot p}_n(t) \,&=\, -\nabla_{q_n}V(q(t))-\hidekappa{\kappa_\R }
(\phi_\R(t),\alpha_\R)_{\HH} ~,\cr 
{\dot \phi}_\L(t) \,&=\,\LL \left(\phi_\L(t) + \hidekappa{\kappa_\L}  \alpha_\L  \cdot
q_1(t)\right) ~,\cr
{\dot \phi}_\R(t) \,&=\, \LL \left(\phi_\R(t) +\hidekappa{ \kappa_\R}  \alpha_\R \cdot
q_n(t)\right)~.
}\EQ(eqmo1)
$$
The last two equations of \equ(eqmo1) are easily integrated and lead to
$$
\eqalign{
\phi_\L(t)\,&=\,\e^{\LL t}\phi_\L{(0)} +\hidekappa{ \kappa_\L}\int_0^t
\!\d s\,\LL e^{\LL(t-s)} \alpha_\L\cdot q_1(s)~, \cr
\phi_\R(t)\,&=\,\e^{\LL t}\phi_\R{(0)} +\hidekappa{ \kappa_\R}\int_0^t
\!\d s\,\LL e^{\LL(t-s)} \alpha_\R\cdot q_n(s)~, \cr
}
$$
where the $\phi_i{(0)}$, $i\in\{\L,\R\}$, are the initial conditions of the
heat baths.
Inserting into the first $2n$ equations of \equ(eqmo1) gives the following
system of integro-differential equations
$$\eqalign{
{\dot q}_j(t) \,&=\, p_j(t)~, \qquad\qquad\qquad j=1,\dots,n ~,\cr
{\dot p}_1(t) \,&=\, -\nabla_{q_1}V(q(t))-\hidekappa{\kappa_\L }
\bigl (\phi_\L{(0)},e^{-\LL t}\alpha_\L \bigr )_{\HH}
-\hidekappa{\kappa_\L^2}\int_0^t \!\d s\, D_\L(t-s)q_1(s) ~,\cr
{\dot p}_j (t) \,&=\, -\nabla_{q_j }V(q(t))~, \qquad\quad j
=2,\dots,n-1 ~,\cr
{\dot p}_n(t) \,&=\, -\nabla_{q_n}V(q(t))-\hidekappa{\kappa_\R }\bigl
(\phi_\R{(0)}, 
e^{-\LL t}\alpha_\R\bigr )_{\HH}
-\hidekappa{\kappa_\R^2}\int_0^t \!\d s\, D_\R(t-s)q_n(s)~, \cr
}\EQ(eqmo2)
$$
where the $d\times d$ dissipation matrices $D_i^{(\mu,\nu)}(t-s)$,
$i\in\{\L,\R\}$, are 
given
by
$$\eqalign{
D_{i}^{(\mu,\nu)}(t-s)\,&=\,
\left(\alpha_{i}^{(\mu)}~,~\LL e^{\LL(t-s)} \alpha_{i}^{(\nu)}
\right)_{\HH}\cr 
             \,&=\,-\frac{1}{d}\delta_{\mu,\nu}\int {\rm dk}\, |{\widehat
\rho_i}(k)|^2
                |k| \sin(|k|(t-s))~.
}$$
The last expression is obtained by observing that 
$$
e^{\LL t}\,=\,\left (\matrix{ \cos(|k|t)& |k|^{-1}\sin(|k|t)\cr
- |k|\sin(|k|t)&\cos(|k|t)}\right )~,
$$
written in Fourier space.

So far we only discussed the finite energy configurations of the heat baths.
We will assume that the two reservoirs are in thermal equilibrium at
inverse temperatures 
$\beta _\L$ and $\beta _\R$. This means that the initial conditions
$\Phi(0) \equiv \{\phi_\L{(0)},\phi_\R{(0)}\}$ 
are distributed according to the Gaussian measure with mean zero and
covariance $\langle \phi_i(f)\phi_j(g)\rangle=\delta_{ij} (1/\beta_i)(f,
g)_{\HH}$. (Recall that the Hamiltonian of the
heat baths is given by $\sum_{i \in \{\L,\R\}}(\phi_i,\phi_i)_{\HH}$.)
If we assume that the coupling functions $\alpha_i^{(\nu)}$ are in
$\HH$, $i \in \{\L,\R\}$, and $\nu \in \{1,\cdots, d \}$  
then the $\xi_i(t) \equiv \phi_i{(0)}(e^{-\LL t}\alpha_i)$
become $d$-dimensional Gaussian random processes with mean zero and covariance
$$
\langle \xi_i(t) \xi_j(s) \rangle\,=\,\delta_{i,j}\frac{1}{\beta_i}
C_i(t-s)~,\quad i,j\in \{\L,\R\}~,
\EQ(cov1)
$$
and the $d\times d$ covariance matrices $C_i(t-s)$ are given by
$$\eqalign{
C_{i}^{(\mu,\nu)}(t-s)
\,&=\,\left(\alpha_{i}^{(\mu)},e^{\LL(t-s)}\alpha_{i}^{(\nu)}\right)_{\HH} \cr 
            \,&=\,\frac{1}{d}\delta_{\mu,\nu}\int {\rm dk}\, |{\widehat
\rho_i}(k)|^2
                 \cos\bigl(|k|(t-s)\bigr)~.
}$$
The relation
$$
{\dot C}_i(t)=D_i(t)~,
\EQ(fluc-diss)
$$
which is checked easily by inspection, is known as the fluctuation
dissipation theorem.
It is characteristic of the Hamiltonian nature of the system.
After these assumptions and transformations,
the equations of motion \equ(eqmo2) become a system of random
integro-differential equations on $\real^{2dn}$ which we will analyze
in the sequel.

Finally, we impose a condition on the random force exerted by the heat baths
on the chain. We assume that 
\medskip
\item{\bf H3)}{The covariances of the random processes $\xi_i(t)$ with
$i\in\{\L,\R\}$
satisfy
$C_i^{(\mu,\nu)}(t-s)\,=\,\delta_{\mu,\nu}\smm \lambda ^2_\im e^{-\gamma_\im |t-s|}$,
with $\gamma_\im>0$
and $\lambda _\im>0$.}
\medskip
This can be achieved by a suitable choice of
the coupling functions $\rho_i(x)$, for example 
$$
\widehat \rho
_i(k)\,=\,\const \prod_{m=1}^M{1\over (k^2 +\gamma_\im^2)^{1/2}}~,
$$
where all the $\gamma_\im$ are distinct.
To keep the notation from still further accumulating, we choose $M$
the same on the left and the right.
We will call the random process
given by \Equ(eqmo2) quasi-Markovian if Condition {\bf H3} is satisfied.
Indeed, using Condition {\bf H3} together with the
fluctuation-dissipation relation \equ(fluc-diss) and enlarging the phase space
one may eliminate the memory terms (both deterministic and random) of
the equations of motion \equ(eqmo2) and rewrite them
as a system of Markovian stochastic differential equations.

By Condition {\bf H3} we can rewrite the stochastic processes $\xi_i(t)$ as
It\^o stochastic integrals
$$
\xi_i(t)=\smm  \lambda _\im\sqrt{\frac{2\gamma_\im}{\beta_i}} \int_{-\infty}^t 
e^{-\gamma_\im (t-s)} \d w_\im(s)~,
$$
where the $ w_\im(s)$ are $d$-dimensional Wiener processes with covariance
$$
{\bf E}\left[\left( w_\im^{(\mu)}(t)-w_\im^{(\mu)}(s)\right) \left(
w_{j,m'}^{(\nu)}(t')-w_{j,m'}^{(\nu)}(s')\right) \right ]
\,=\, \delta_{i,j} \delta_{\mu,\nu} \delta_{m,m'} 
\vert [s,t]\cap [s',t'] \vert~,
\EQ(exp) 
$$
where  $s < t$ and $s' < t'$,  ${\bf E}$ is the expectation on the
probability space of the Wiener process 
and $\vert\, \cdot \vert$ denotes the
Lebesgue measure.
We introduce new ``effective'' variables $r_\Lm$, $r_\Rm  \in \real^d
$, with $\mlist$,
which describe both the retro-action of the heat bath onto the system
and the random force exerted by the heat baths:
$$
\eqalign{
r_\Lm(t)\,&=\,\hidekappa{ \kappa_\L^2}  \lambda ^2_\Lm\gamma_\Lm  \int_0^t \d
s\, e^{-\gamma_\Lm  (t-s)} q_1(s) 
     ~- \hidekappa{\kappa_\L } \lambda _\Lm \sqrt{\frac{2\gamma_\Lm}{\beta_\L}}
\int_{-\infty}^t e^{-\gamma_\Lm (t-s)} \d w_\Lm(s)~, \cr
r_\Rm (t)\,&=\,\hidekappa{ \kappa_\R^2 } \lambda^2 _\Rm\gamma_\Rm
\int_0^t \d s\, e^{-\gamma_\Rm  (t-s)} q_n(s) 
   ~-\hidekappa{ \kappa_\R } \lambda _\Rm \sqrt{\frac{2\gamma_\Rm}{\beta_\R}} 
\int_{-\infty}^t e^{-\gamma_\Rm (t-s)} \d w_\Rm (s) ~. 
}$$
We get the following system of Markovian stochastic
differential equations
$$\eqalign{
\d q_j (t) \,&=\, p_j (t)\dt~, \qquad\qquad\qquad j =1,\dots,n ~,\cr
\d p_1 (t) \,&=\, -\nabla_{q_1}V(q(t))\dt  +\smm r_\Lm(t)\dt~,\cr
\d p_j (t) \,&=\, -\nabla_{q_j }V(q(t))\dt ~,\qquad j =2,\dots,n-1 {}~,\cr
\d p_n (t) \,&=\, -\nabla_{q_n}V(q(t))\dt  +\smm  r_\Rm (t)\dt ~,\cr
\d r_\Lm (t) \,&=\, -\gamma_\Lm  r_\Lm(t)\dt + \lambda_\Lm^2
\gamma_\Lm q_1(t) \dt
                 - \lambda_\Lm\sqrt{\frac{2\gamma_\Lm}{\beta_\L}}\, \d w_\Lm(t)  ~,\cr
\d r_\Rm (t) \,&=\, -\gamma_\Rm  r_\Rm (t)\dt + \lambda_\Rm^2
\gamma_\Rm q_n(t) \dt 
                 - \lambda_\Rm \sqrt{\frac{2\gamma_\Rm}{\beta_\R}}\, \d w_\Rm(t)~,  \cr
}\EQ(eqmo3)
$$
which defines a Markov diffusion process on $\real^{2d(n+M)}$.
This system of equations is our main object of study.
Our main results are the following:

\THEOREM(1) If Conditions {\bf H1}-{\bf H2} hold,
there is a constant $\lambda^* > 0$,
such that for $|\lambda_\Lm|$, $|\lambda_\Rm| \in(0,
\lambda^*)$ with $\mlist$, the solution of
\Equ(eqmo3) is a Markov process
which has an absolutely continuous invariant measure
$\mu$ with a $\CC^{\infty}$ density.

\REMARK In \clm(I.posnew) we will show even a little more. Let
$h_0(\beta )$ be the Gibbs distribution for our system when the 
heat baths are both at temperature $1/\beta $. If $h$ denotes the
density of the
invariant measure found in \clm(1), we find that $h/h_0(\beta ) $ is
in the Schwartz space $\SS$ for all
$\beta<\min(\beta_\L,\beta_\R)$.
This mathematical statement reflects the intuitively obvious fact that
the chain can not get hotter than either of the baths.

Concerning the uniqueness and the ergodic properties of the invariant
measure, our results are restricted to small temperature differences.
We have the following result.

\THEOREM(2) If Conditions {\bf H1}-{\bf H2} hold, there are constants 
$\lambda^* > 0$ and $\epsilon>0$
such that for $|\lambda_\Lm|$, $|\lambda_\Rm| \in (0,
\lambda^*)$ with $\mlist$, and
$|\beta_\L-\beta_\R|/(\beta_\L+\beta_\R) < \epsilon$, the Markov
process \equ(eqmo3) has a unique invariant measure and this measure is
mixing.

\REMARK The restriction on the couplings between the small system and
the baths $\lambda_\Lm$, $\lambda _\Rm$
is non-perturbative: it is a condition of stability of
the coupled system small system plus heat baths. 
Indeed, the baths have the effect of renormalizing the  deterministic potential
seen by the small system. The constant $\lambda^*$ depends only
on the potential $V(q)$: if the coupling constants
$\lambda_\Lm$, $\lambda _\Rm$  are too large, the
effective potential ceases to be stable and, at least at equilibrium
({\it i.e.}, for $\beta_\L=\beta_\R$), there is no invariant
probability measure for the Markov process \equ(eqmo3), but
only a $\sigma$-finite invariant measure (see \Equ(effective) and \Equ(hh0)).
This restriction is related to Condition {\bf H1} on the potential:
for potentials which grow at infinity faster than quadratically, this
restriction would not be present (see \CITE[JP1]).
On the other hand, the restriction on the temperature difference {\em
is} of perturbative origin.

A more physical formulation of the results of \clm(2) is obtained by
going back to  the Eq.\equ(eqmo2), which expresses all the quantities
in terms of the  phase space of the small system and the initial
conditions $\Phi(0)$
of the heat baths. 
Let us introduce some notation: For given initial conditions
$\Phi(0)$, we let 
$\Theta_{t,\Phi(0)}(p,q)$ denote the solution of Eq.\equ(eqmo2).
Finally, define
$$
\nu(\dpp,\dq)\,=\,\int_{r\in\real^{2dM}} \mu(\dpp,\dq,\dr)~,
$$
where $\mu$ is the invariant measure of \clm(1).

\CLAIM Corollary(c1) Under the hypotheses of
\clm(2), the system Eq.\equ(eqmo2) reaches a stationary state and is
mixing in the following sense: For any observables $F$, $G
\in \L^2({\bf R}^{2dn},\nu(\dpp,\dq))$ and for any probability measure
$\nu_0(\dpp,\dq)$ which is absolutely continuous with respect to
$\nu(\dpp,\dq)$ we have
$$
\eqalign{
\lim_{t\to\infty } \int\!\! \nu_0(\dpp,\dq) \bigl\langle \bigl(F\circ
\Theta_{t,\Phi_0}\bigr)(p,q) \bigr\rangle \,&=\, \int\!\!
\nu(\dpp,\dq) F(p,q)~,\cr 
\lim_{t\to\infty } \int \nu(\dpp,\dq)
\big\langle  
\bigl( F\circ \Theta_{t,\Phi_0}\bigr )(p,q)
G(p,q)\big\rangle \,&=\, 
\int\!\! \nu(\dpp,\dq) F(p,q) \int\!\! \nu(\dpp,\dq) G(p,q)~.\cr
}
\EQ(limit1)
$$
Here, $\langle\cdot\rangle$ denotes the integration over the Gaussian
measures of the two heat baths, introduced earlier.

We explain next the strategy to prove these results.
Our proof is based on a detailed study of Eq.\equ(eqmo3). Let $x=(p,q,r)$.
For a Markov process $x(t)$ with phase space $X$
and an invariant measure $\mu({\dx})$, its ergodic
properties may be deduced from the study of the associated semi-group $T^t$
on the Hilbert space $\L ^2(X,\mu({\dx}))$. To prove the existence
of the invariant measure in \clm(1) we proceed as follows: We
consider first 
the semi-group $T^t$ on the auxiliary Hilbert space $\HH_0\equiv\L
^2(X,\mu_0({\dx}))$, 
where the reference measure $\mu_0({\dx})$ is a generalized Gibbs
state for a suitably chosen reference temperature. Our main technical result
consists in proving that the generator $L$ 
of the semi-group $T^t$ on $\HH_0$ and its
adjoint $L^{*}$ have compact resolvent.
This is proved by generalizing the commutator method developed by
H\"ormander to study hypoelliptic operators. From this follows the existence
of a solution to the eigenvalue equation $(T^t)^{*}g=g$ in
$\HH_0$ and this implies immediately the existence of
an invariant measure. To prove \clm(2) we use a perturbation
argument, indeed at equilibrium ({\it i.e.}, for $\beta_\L=\beta_\R$) the
invariant measure is unique and $0$ is a simple eigenvalue of the
generator $L$ in $\HH_0$. Using the compactness
properties of $L$, we show that $0$ is a simple eigenvalue of the
generator $L$ in $\HH_0$ for
$|\beta_\L-\beta_\R|/(\beta_\L+\beta_\R)$ 
small enough. And this can be used to prove the uniqueness claim of 
\clm(2), while the mixing properties will be shown by extending 
the method of \CITE[Tr].

This paper is organized as follows: In Section 3 we prove \clm(1) and
\clm(2) except for our main estimates \clm(I.com) and
\clm(schw) which are proven in Section 4. In Appendices A, B and C,
we prove some auxiliary results.

%
\section{Invariant Measure: Existence and Ergodic Properties} 
%

In this section, our main aim is to prove  \Theorem(1) and
\Theorem(2). We first prove some basic consequences of our
Assumptions {\bf H1} and {\bf H2}. 
In particular, we define the semi-group $T^t$
describing the solutions of \Equ(eqmo3) on the auxiliary Hilbert
space $\HH_0$ described in the introduction. Furthermore we recall
some basic facts on hypoelliptic differential operators. Once these
preliminaries are in place, we can attack the proof of \Theorem(1) and
\Theorem(2) proper.

\SUBSECTION Existence and fundamental properties of the dynamics

Let $X=\real^{2d(n+M)}$ and write the stochastic differential
equation \equ(eqmo3) in the abbreviated form
$$
dx(t)\,=\,b(x(t))+\sigma dw(t)~,
\EQ(sde)
$$
where
\item{(i)} $b$ is a $\CC^{\infty}$ vector field which satisfies, by
Condition {\bf H1},
$$
\sup_{x\in X} |\pa^{\alpha} b(x)|\,<\,\infty~,
$$
for any multi-index $\alpha$ such that $|\alpha|\ge 1$. In particular
$$
B\equiv \|{\rm div~} b\|_{\infty}\,<\,\infty ~.
\EQ(div)
$$
\item{(ii)} $\sigma: \real^{2dM}\rightarrow X$ is a linear map. We also
define
$$
D\,\equiv\,\frac{1}{2} \sigma \sigma^T \,\ge\, 0~.
\EQ(thed)
$$
\item{(iii)} $w \in \WW \equiv \CC(\real;\real^{2dM})$ is a standard
$2dM$-dimensional Wiener process. 

The Eq.\equ(sde) is rewritten more precisely as the integral equation
$$
\xi(t,w;x)\,=\,x+\int_0^t\!ds\, b(\xi(s,w;x)) + \sigma(w(t)-w(0))~.
\EQ(sol)
$$
It follows from an elementary contraction argument (see e.g. [Ne], Thm
8.1) that \equ(sol) has a unique solution 
$$
\real \ni t \,\mapsto\, x(t)= \xi(t,w;x) \in \CC(\real;X)~,
$$
for arbitrary initial condition $x\in X$ and $w\in \WW$.

The difference $w(t)-w(0)$ has the statistics of a standard Brownian motion
and we denote by ${\bf E}[\cdot]$ the corresponding expectation. By well-known
results on stochastic differential equations, this induces on
$\xi(t,w;x)$ the statistics of a Markovian diffusion process with
generator
$$
\nabla\cdot D \nabla + b(x)\cdot\nabla~.
\EQ(gena)
$$
More precisely (see [Ne] Theorem 8.1): Let
$\CC_{\infty}(X)$ denote the continuous functions which vanish at
infinity with the sup-norm and
let $\FF^t$ be
the $\sigma$-field generated by $x$ and $\{w(s)-w(0)\,;\,0< s \le t\}$,
then for $0 \le s \le t$ and $f \in \CC_{\infty}(X)$ we have
$$
{\bf E}\left[ f(x(t)) \vert \FF^s \right]\,=\, T^{t-s}f(x(s)) \qquad
{\rm ~a.s.}~,
\EQ(TT)
$$
where $T^t$ is a strongly continuous contraction semi-group of
positivity preserving operators on $\CC_{\infty}(X)$ whose
generator reduces to \equ(gena) on $\CC^{\infty}_{{ 0 }}(X)$.

In the sequel we denote by $L$ the differential operator \equ(gena)
with domain $D(L)=\CC^{\infty}_{{ 0 }}(X)$.

To prove the existence of an invariant measure we will study the
semi-group $T^t$ or rather an extension of it 
on the auxiliary weighted Hilbert space 
$\HH_0$ described in the introduction.
To define $\HH_0$ precisely, we consider the ``effective Hamiltonian''
$$
G(p,q,r)\,=\,H_\S(p,q)+\smm\biggl(\frac{1
}{\lambda_\Lm^2}\frac{r_\Lm^2}{2} 
 + \frac{1 }{\lambda_\Rm^2}\frac{r_\Rm^2}{2}
- q_1\cdot r_\Lm  
            - q_n\cdot r_\Rm \biggr) ~.
\EQ(effective)
$$
We note that, due to Condition {\bf H1}, $G(x) \rightarrow +\infty$ as
$|x| \rightarrow \infty$ as long as $|\lambda_\Lm|$, $|\lambda_\Rm|<
\lambda^*$ for some $\lambda^*$ depending only on the potential
$V(q)$.

We choose further a ``reference temperature'' $\beta_0$, which is
arbitrary subject to the condition 
$$
\beta _0\,<\,2\min(\beta _\L,\beta _\R)~.
\EQ(beta)
$$ 
For example we could take $\beta_0$ as the inverse of the mean temperature of
the heat baths: $\beta_0^{-1}=(\beta_\L^{-1}+\beta_\R^{-1})/2$. For
the time being,
it will be convenient not to fix $\beta_0$.
Then, we let
$$
\HH_0\,=\,
\L ^2(X,Z_0^{-1}e^{-\beta_0 G}\,{\dx})~,
\EQ(hh0)
$$
and we denote $(\cdot , \cdot)_{\HH_0}$ and $\|\cdot\|_{\HH_0}$ the
corresponding scalar product and norm.

\REMARK With a proper choice of $Z_0$, it is easy to check that the quantity
$Z_0^{-1}e^{-\beta_0 G(q,p,r)}\,{\dx}$ is the
invariant measure for the Markov process \Equ(eqmo3) when $\beta
_\L=\beta _\R=\beta _0$ and $|\lambda_\Lm|, |\lambda_\Rm| <
\lambda^*$.

\LEMMA(I.1) If the potential $V$ satisfies Condition {\bf H1} and if 
$\beta_0 < 2\min(\beta_\L,\beta_\R)$  there is a $\lambda^* > 0$ such
that if the couplings satisfy $|\lambda_\Lm|$, $|\lambda_\Rm|\in 
(0,\lambda^*)$, then
the semi-group $T^t$ given by \Equ(TT) extends to
a strongly continuous quasi bounded semi-group $\TTH0$ on $\HH_0$:
$$
\|\TTH0\|_{\HH_0}\,\le\,e^{\alpha t}~,
$$
where $\alpha$ is given by
$$
\alpha \,=\, d \smm \left (\gamma_\Lm \left ({1\over 2}-{\sqrt{(\beta
_\L-\beta_0/2)\beta_0/2}\over \beta _\L}\right )
+ \gamma_\Rm \left ({1\over 2}-{\sqrt{(\beta
_\R-\beta_0/2)\beta_0/2}\over \beta _\R}\right )\right )
\,\ge\,0~.
\EQ(alpha)
$$ 
The generator $L_{\HH_0}$ of $\TTH0$ is the closure of the differential
operator $L$ with domain $\CC^{\infty}_{{ 0 }}$ given by
$$\eqalign{
L\,&=\,\smm{\lambda_\Lm^2 \gamma_\Lm\over \beta_\L}
\left (\nabla_{r_\Lm}-\beta_\L \zl\right ) 
\cdot\nabla_{r_\Lm }\cr 
\,&+\,\smm{\lambda_\Rm^2 \gamma_\Rm\over \beta_\R}
\left (\nabla_{r_\Rm}-\beta_\R \zr\right ) 
\cdot\nabla_{r_\Rm } \cr
  \,&+\, \smm r_\Lm \cdot\nabla_{p_1} + L_\S  +  
\smm r_\Rm \cdot\nabla_{p_n}~, \cr
}
\EQ(gen1)$$
with the abbreviations
$$
\zl\,=\,\lambda_\Lm^{-2}r_\Lm-q_1~,\quad
\zr\,=\,\lambda_\Rm^{-2}r_\Rm-q_n~,\quad
\EQ(zetadef)
$$
and where
$L_\S $ is the Liouville operator associated with the Hamiltonian
$H_\S(q,p)$: 
$$
L_\S \,=\,\sum_{j=1}^{n} p_j\cdot\nabla_{q_j} - 
(\nabla_{q_j}V)\cdot\nabla_{p_j}~.
\EQ(Liouv)
$$
Moreover, $\TTH0$ is positivity preserving:
$$
\TTH0 f \,\ge\, 0 \quad {\rm if}\quad  f\,\ge\, 0~,
\EQ(posi)
$$
and 
$$
\TTH0 1\,=\,1~.
\EQ(one)
$$

\REMARK We have $\alpha =0$ if only if $\beta _\L=\beta _\R=\beta_0$. 

\PROOF The proof uses standard tools of stochastic analysis
and is given in Appendix A.

Having shown a priori bounds using Condition {\bf H1}, we will state
one basic consequence of Condition {\bf H2}. We recall that a
differential operator $P$ is called hypoelliptic if
$$
{\rm sing~}{\rm supp}~ u\,=\,{\rm sing~}{\rm supp }~Pu \quad {\rm
for~all~} u \in \DD'(X)~.
$$
Here $\DD'(X)$ is the usual space of distributions on the infinitely
differentiable functions with compact support and for $u \in \DD'(X)$,
${\rm sing~}{\rm supp}~u$ is the set of points $x\in X$ such that
there is no open neighborhood of $x$ to which the restriction of
$u$ is a $\CC^{\infty}$ function.

Let $P$ be of the form 
$$
P\,=\,\sum_{j=1}^{J} Y_j^2 + Y_0~,
\EQ(Kol)
$$ 
where
$Y_j$, $j\in\{0,\dots,J\}$ are real $\CC^{\infty}$ vector fields. Then
by H\"ormander's Theorem, [H\"o], Thm 22.2.1, if the Lie algebra
generated by $Y_j$, $j\in\{0,\dots,J\}$ has rank dim $X$ at every
point, then $P$ is hypoelliptic.

Differential operators arising from diffusion problems are of the form
\equ(Kol). Let $L$ be the differential operators given in \Equ(gen1),
let $L^T$ denote its formal adjoint, then one may easily check that
Condition {\bf H2} implies that any of the following operators
$$
L, \quad L^T, \quad \pa_t + L, \quad  \pa_t + L^T~,
$$
satisfies the condition of H\"ormander's Theorem and thus is
hypoelliptic.
As an immediate consequence we have:
\CLAIM Corollary(smef) If Condition {\bf H2} is
satisfied then the 
eigenvectors of $L$ and $L^T$ are $\CC^{\infty}$ functions.

Next, let $P(t,x,E)$, $t\ge 0$, $x\in X$, $ E \in \BB$ denote the transition
probabilities of the Markov process $\xi(t,w;x)$ solving the
stochastic differential equation \equ(eqmo3) with initial condition
$x$, {\it i.e.},
$$
P(t,x,E)\,=\,{\bf P}\left( \xi(t,w:x) \in E \right)~, 
$$
where ${\bf P}$ denotes the probability associated with the Wiener
process. Then by the forward and backward Kolmogorov equations
we obtain 
\CLAIM Corollary(trprob) If Conditions {\bf H1} and {\bf H2} are
satisfied then the transition 
probabilities of the Markov Process $\xi(t,w;x)$ have a smooth density
$$
P(t,x,y)\,\in \, \CC^{\infty}((0,\infty)\times X \times X)~.
$$

\SUBSECTION Proof of \clm(1) and \clm(2)

After these preliminaries we
now turn to the study of spectral properties of the generator
$L_{\HH_0}$ of the semi-group $\TTH0$.

The proof of the existence of the invariant measure will 
be a consequence of the following key property which we prove in Section 4.
\PROPOSITION(I.com) If the potential $V$ satisfies Conditions {\bf
H1}, {\bf H2} and if $\beta_0 < 2\min(\beta_\L,\beta_\R)$ there is a
$\lambda^*>0$ such that if the couplings  satisfy $|\lambda_\Lm|$,
$|\lambda_\Rm|\in (0,\lambda^*)$
then both $L_{\HH_0}$ and $L_{\HH_0}^{*}$ have compact resolvent.

A useful by-product of the proof \clm(I.com) are some additional
smoothness and decay properties of the
eigenvalues of $L_{\HH_0}$ and $L_{\HH_0}^{*}$ on $\HH_0$.

\PROPOSITION(schw) Let $g$ denote an eigenvector of $L_{\HH_0}$ or
$L_{\HH_0}^{*}$. If the assumptions of \clm(I.com) are satisfied 
then we have
$$
g e^{-\beta_0 G/2}\,\in \SS(X)~,
$$ 
where $\SS(X)$ denotes the Schwartz space.

Using these results, we come back to the Markov process defined by the
Eqs.\equ(eqmo3), and whose semi-group $T^t$ was defined in
Eq.\equ(TT). We prove the existence of an invariant measure with a
smooth density and give a bound which shows that, in some sense,
the chain does not get hotter than the hottest heat bath.

\PROPOSITION(I.posnew) Under the assumptions {\bf H1}--{\bf H2} there
is a $\lambda^*>0$ such that if the couplings  satisfy
$|\lambda_\Lm|$, $|\lambda_\Rm| \in (0,\lambda^*)$ the Markov process $T^t$
has an invariant measure $\mu$ which is
absolutely continuous with respect to
the Lebesgue measure. Its density $h$ satisfies the following: $h
\exp(\beta G) \in \SS(X)$ for all $\beta < \min(\beta_\L,\beta_\R)$.

\PROOF The function 1 is obviously a solution of $L f = 0$ with $L$
defined in Eq.\equ(gen1). Note next that the function 1 is in $\HH_0$,
as is seen from Eq.\equ(hh0) (if $|\lambda_\Lm|$ and $|\lambda_\Rm|$
are sufficiently small). Since, by \clm(I.com), the operator $L_{\HH_0}$ has
compact resolvent on $\HH_0$, it follows that
$0$ is also an eigenvalue of $L_{\HH_0}^*$. Let us denote the corresponding
eigenvector by $g$. We will choose the normalization $(g,1)_{\HH_0}=1$. 
We assume first that $g\ge0$. Then the function
$$
h(x)\,=\, Z_0^{-1} g(x) e ^{-\beta _0 G(x)}~,
\EQ(im)$$
with $\beta _0$ and $G$ defined in Eqs.\equ(beta) and \equ(effective),
is the density of an invariant measure for the process $T^t$:
Indeed,
we note first that $\|h\|_{\L ^1(X,{\dx})}=(1,g)_{\HH_0}$ is
finite and thus 
$\mu({\dx})$ is a probability measure. 

Let $E$ be some Borel set. Then the characteristic function $\chi_E$ of $E$ 
belongs to $\HH_0$. We have
$$\eqalign{
\int \mu({\dx})\, T^t \chi_E \,&=\, Z_0^{-1} \int\!\!\dx\, e ^{-\beta _0 G(x)}
g(x) T^t \chi_E \cr 
\,&=\,Z_0^{-1} \int\!\!\dx\, e^{-\beta _0 G(x)}
(\TTH0)^{*}g(x)\chi_E \cr
\,&=\, \mu(E)~,  
}$$
and therefore $\mu({\dx})$ is an invariant measure for the Markov
process \equ(eqmo3).

To complete the first part of the 
proof of \clm(I.posnew) it remains to show that
$g\ge0$. We will do this by checking that $h\ge0$.
We need some notation. Let $L^\T$ denote the formal adjoint of
$L$. Then one has $L^\T h=0$. This follows from the identities
$$
\eqalign{
\int\! \dx\, f L^\T h\,&=\, Z_0^{-1} \int\! \dx\, f L^\T \bigl (g e^{-\beta
_0G}\bigr )\cr
\,&=\, Z_0^{-1} \int\! \dx\, \bigl (L f) g e^{-\beta
_0G}\cr
\,&=\, \bigl (Lf,g)_{\HH_0}\,=\,(f,L_{\HH_0}^*g
)_{\HH_0}\,=\,(f,0)_{\HH_0}\,=\,0~,
\cr
}
$$
which hold for all $f\in \CC^{\infty}_{{\rm 0}}(X)$.
Consider now the semi-group $T^t$ acting on the space $\CC_{\infty}(X)$
defined at the beginning of Section 3. 
Since $T^t$ is a Markovian semi-group, we have $T^t 1=1$ and $T^t$ is
positivity preserving.
The operator $T^t$ induces an action $(T^t)^*$
defined on the dual space $\CC^*_{\infty}(X)$ which consists of finite measures.
Since $T^t$ is Markovian, $(T^t)^*$ maps probability measures to
probability measures.
Furthermore, if a measure $\nu$ has a density $f$ in $\L ^1(X,\dx)$, then
$(T^t)^* \nu$ is a measure which has again a density in $\L ^1(X,\dx)$: 
Indeed, by \clm(trprob)
the transition probabilities of the Markov process $P(t,x,y)$ are in
$\CC^{\infty}\left((0,\infty)\times X \times X \right) $. 
If we denote by $(T^t)^\T$
the induced action of $(T^t)^*$ on the densities, we have for $g\ge0$,
$$
\eqalign{
(T^t)^* \bigl (g(x)\dx\bigr )\,&=\,\int {\rm dy}\,
g(y) P(t,y,\dx)\,=\,\dx \int {\rm dy}\,
g(y) P(t,y,x)\,=\,\bigl ((T^t)^\T g \bigr )(x)\dx~,
}
$$
and  $\|(T^t)^\T g \|_{\L ^1}=\|g\|_{\L ^1}$.

Coming back to the invariant density $h$, we know that 
$$
(T^t)^\T h\,=\,h~.
$$
We next show $(T^t)^\T |h|=|h|$. Since $|h|\pm h\ge 0$, we have
$(T^t)^\T (|h|\pm h)\ge 0$. This can be rewritten as
$$
\bigl |(T^t)^\T h\bigr |\,\le\,(T^t)^\T |h|~.
$$
Therefore,
$$
|h|\,=\,\bigl |(T^t)^\T h\bigr |\,\le\,(T^t)^\T |h|~.
$$
Since $(T^t)^\T$ preserves the $\L ^1$-norm, we conclude that
$$
|h|\,=\,(T^t)^\T |h|~.
\EQ(ouff)
$$
This shows the existence of an invariant measure.

Now, by \clm(schw), we have $h \exp(\beta G/2)\in\SS(X)$ for all $\beta <
2\min (\beta_\L,\beta_\R)$ and so for $\beta < \min
(\beta_\L,\beta_\R)$ it follows that
$h \exp(\beta G)\in\SS(X)$. This concludes the proof of \clm(I.posnew).

We next prove the uniqueness of the invariant measure and the ergodic
properties of the Markov process. We start by fixing an inverse
temperature $\beta_0$. If $\beta _\L=\beta _\R =\beta _0$, the two
heat baths are at the same temperature, and the equilibrium state of
the system is known, since it is given by the generalized Gibbs distribution
$Z_0^{-1}e^{-\beta _0 G}$. For the equilibrium case, this distribution
is the unique invariant measure. The existence is obvious from what we
showed for the case of arbitrary temperatures. To show uniqueness,
assume that there is a second invariant measure. Since $L^T$ is
hypoelliptic, then by \clm(smef) this measure has a smooth
density. Since different smooth invariant measures have mutually
disjoint supports and $e^{-\beta _0 G}$ has support everywhere, uniqueness
follows. If the invariant measure is unique, it is ergodic and hence,
(see [Yo] and [Ho]) $0$ is a simple eigenvalue of $L_{\HH_0}$.

The case of different temperatures will be handled by a perturbation
argument around the equilibrium situation we just described. This perturbation
argument will take place in the {\em fixed} Hilbert space $\HH_0$
defined in Eq.\equ(hh0). Thus, we will consider values of $\beta _\L$
and $\beta _\R$ such that
$$
{1\over \beta _0}\,=\,\frac{1}{2} \bigl ( {1\over \beta _\L} + {1\over \beta
_\R}\bigr )~,\qquad
\frac{|\beta_\L-\beta_\R|}{\beta_\L+\beta_\R}\,<\,\epsilon~,
\EQ(conditions)
$$
for some small $\epsilon > 0$ (which does not depend on $\beta _0$).

We first show that $0$ remains a simple eigenvalue
of the generator $L_{\HH_0}$ when the temperature difference
satisfies \equ(conditions) for a sufficiently small $\epsilon$.


\CLAIM Lemma(simplicity) Under the assumptions {\bf H1}--{\bf H2} there
are constants  $\lambda^*>0$ and $\varepsilon >0$
such that if the couplings  satisfy $|\lambda_\Lm|$, $|\lambda_\Rm|\in
(0,
\lambda^*)$  
and moreover $\beta_\L$, $\beta _\R$ satisfy \equ(conditions),
then  $0$ is a simple eigenvalue of the generator $L_{\HH_0}$.

\PROOF It will be convenient to work in the flat Hilbert space $\L
^2(X, {\dx})$. Note that $\Calkop=\exp{(-\beta_0 G /2)} L \exp{(+\beta_0
G /2)}$ is a function 
$\Calkop\equiv\Calkop(\beta _\L,\beta _\R,\beta _0)$. We write this as
$$
\Calkop (\beta_\L ,\beta_\R,\beta _0)\,=\,
\Calkop (\beta_0,\beta_0,\beta _0) + \delta Z~,
$$
where
$$
\delta\,=\,\frac{\beta_\R-\beta_\L}{\beta_\R+\beta_\L}~.
$$
One finds
$$
\eqalign{ 
\Calkop (\beta_0,\beta_0,\beta _0)\,&=\, \smm\frac{\lambda_\Lm^2\gamma_\Lm}{2}\left(
   \frac{2}{\beta_0}\nabla_{r_\Lm }^2-\frac{\beta_0}{2}
   \zl ^2 +
    \frac{d}{\lambda_\Lm^2}\right)\cr 
                                  &~+\smm\frac{\lambda_\Rm^2\gamma_\Rm}{2}\left( 
   \frac{2}{\beta_0}\nabla_{r_\Rm }^2-\frac{\beta_0}{2}
   \zr ^2 +
    \frac{d}{\lambda_\Rm^2}\right)  \cr
&~+ \smm r_\Lm \cdot\nabla_{p_1} + L_\S  + \smm r_\Rm \cdot\nabla_{p_n}~,
}$$
and
$$\eqalign{
Z\,&=\,\smm\frac{\lambda_\Lm^2\gamma_\Lm}{2}
    \biggl(\frac{2}{\beta_0}\nabla_{r_\Lm }^2  
        +
\zl \cdot\nabla_{r_\Lm } 
        + \nabla_{r_\Lm }\cdot \zl 
        +\frac{\beta_0}{2}\zl ^2 
         \biggl)\cr     
 ~~\,&-\,\smm\frac{\lambda_\Rm^2\gamma_\Rm}{2}
    \biggl(\frac{2}{\beta_0}\nabla_{r_\Rm }^2  
        +
        \zr \cdot\nabla_{r_\Rm }  
        + \nabla_{r_\Rm }\cdot \zr 
        +\frac{\beta_0}{2}\zr ^2
         \biggr) ~. 
}
$$
Furthermore, by \Proposition(I.com), 
$R_0\equiv(1-\Calkop (\beta_0,\beta_0,\beta _0))^{-1}$ is a compact
operator, and therefore the simple eigenvalue 1 of $R_0$ is isolated.
From now on we assume for convenience
that $\alpha\equiv\alpha(\beta_\L ,\beta_\R)$ is strictly 
smaller than one. Note that this is no restriction of generality: 
if $\alpha
\in[n-1, n)$ with $n>1$, we replace $(1-\Calkop )^{-1}$ by
$(1-\frac{1}{n}\Calkop )^{-1}$ in the following 
discussion.  

We show next that the resolvent 
$R(\beta_\L ,\beta_\R,\beta _0)\equiv(1-\Calkop (\beta_\L ,\beta_\R,\beta _0))^{-1}$  
depends analytically on the parameter $\delta$.
It is convenient to
write the perturbation $Z$ as
$$
Z\,=\,\sum_{j=1}^{N} E_j F_j~,
$$
where the $E_j$ and $F_j$ are of the form
$\const \partial_{r^{(\nu)}_\im}$ or $\const \zi^{(\nu)}$, $i \in
\{\L,\R\}$, $\mlist$, and $N=8 d M$.
With the matrix notation 
$$
F\,=\,\left(\matrix {F_1 \cr \vdots \cr F_{N}\cr }\right)~,\qquad
E^T\,=\,\left(E_1, \ldots, E_{N}\right)~,
$$
we can write $Z$ as $Z=E^TF$. We will use the following resolvent formula 
$$
R(\beta_\L,\beta_\R,\beta_0)\,=\,R_0 \left(1 + \delta R_0 E^T 
\left(1- \delta F R_0 E^T\right)^{-1} F R_0\right) 
~.
\EQ(resol)$$
To justify \Equ(resol) we have to show that for $\delta$
small enough the operator-valued matrix 
$\left(1-\delta FR_0E^T\right)$ is invertible. It is enough to show
that $F_jR_0E_k$ 
is a bounded operator, for all $j,k$. For this we decompose
$1-\Calkop (\beta_0,\beta_0,\beta _0)$ into its symmetric and antisymmetric parts:
$$
1-\Calkop (\beta_0,\beta_0,\beta _0)\,=\,X+iY~,
$$
where
$$\eqalign{
X\,=\,1\,&+\,\smm\frac{\lambda_\Lm^2\gamma_\Lm}{2}\left(-
   \frac{2}{\beta_0}\nabla_{r_\Lm }^2\,+\,\frac{\beta_0}{2}
   \zl ^2 - 
        d\frac{1}{\lambda_\Lm^2}\right) \cr
  \,&+\,\smm\frac{\lambda_\Rm^2\gamma_\Rm}{2}\left(- 
   \frac{2}{\beta_0}\nabla_{r_\Rm }^2+\frac{\beta_0}{2}
   \zr ^2 - 
       d \frac{1}{\lambda_\Rm^2}\right)~. \cr
}
$$
From the simple estimates
$$\eqalign{
\|E_jf\|^2\,&\le\, (f,Xf)~,  \cr
\|F_jf\|^2\,&\le\,(f,Xf) ~, 
}$$
which hold for $f\in \CC^{\infty}_{\rm 0}(X)$, and since $X$ is a strictly
positive operator we see 
that $E_jX^{-1/2}$ and $F_jX^{-1/2}$ are bounded operators. From the identity
$$\eqalign{
E_j R_0 F_k 
      \,&=\, E_j (X+iY)^{-1} F_k \cr
      \,&=\, E_j X^{-1/2} (1+iX^{-1/2}YX^{-1/2})^{-1} X^{-1/2} F_k~, \cr
}$$
we see that $E_j R_0 F_k$ are bounded operators for all $j,k$. 
Therefore the r.h.s of \Equ(resol) is well defined for sufficiently
small $\delta$.
An immediate consequence of the resolvent formula \equ(resol) is that for
sufficiently small $\delta$ the spectrum of
$R(\beta_\L ,\beta_\R,\beta_0)$  has the same form  as the spectrum
$R_0$: $1$ is an eigenvalue and there
is a spectral gap and, in particular $1$ is a simple eigenvalue.
This concludes the proof of \clm(simplicity).

Next we use this lemma to prove
uniqueness of the invariant measure.
We have the following
\CLAIM Theorem(uniqueness) Under the assumptions {\bf H1}--{\bf H2} there
are constants  $\lambda^*>0$ and $\varepsilon >0$
such that if the couplings  satisfy $|\lambda_\Lm|$, $|\lambda_\Rm|\in
(0,\lambda^*)$  
and the temperatures satisfy
$|\beta_\R-\beta_\L|/(\beta_\L+\beta_\R)<\epsilon$, 
the Markov process $T^t$ has a unique (and hence ergodic) invariant 
measure. 

\PROOF The proof uses a dynamical argument.
By \clm(I.com) we have in the Hilbert space $\HH_0$ (with $\beta _0$
given as in \equ(conditions)) the
eigenvalue equation $\TTH0 1=1$ and $(\TTH0)^{*}g=g$. Let the eigenvectors be
normalized such that $(g,1)_{\HH_0}=1$. By \clm(simplicity), $0$
is a simple eigenvalue of the generator $L_{\HH_0}$ if 
$(\beta_\R-\beta_\L)/(\beta_\L+\beta_\R)$ is small enough
and by \clm(I.posnew), the measure $\mu(\dx)=Z_0^{-1}g\exp{(-\beta_0G)}$ is an
invariant measure for the Markov  
process. It is absolutely continuous with respect to the Lebesgue
measure which we denote by $\lambda$.

Assume now that $\nu$ is another invariant probability measure. By
the hypoellipticity of $L$ it must have a smooth density. Therefore
there is a Borel set $A\subset X$, which we may assume bounded, with the
following properties: we have $\nu(A)>0$ and $\lambda(A)>0$ but
$\mu(A)=0$, because the measures have disjoint supports.
Let $\chi_A$ denote the characteristic function of the set
$A$. By the pointwise ergodic theorem, see \CITE[Yo] and \CITE[Ho], we
have, denoting $\sigma_s(x)$ 
the ergodic average $\sigma_s(x)=(1/s)\int_0^s \dt\, T^t \chi_A(x)$,
$$
\lim_{s\rightarrow\infty} \sigma_s(x)\,=\,\nu(A)~, \quad \nu{\rm\hbox{\rm--a.e.}}~.
\EQ(birerg)
$$
Since $T^t$ is a contraction semi-group on $B(X,{\mathcal{B}})$, and
$\chi_A\le1$ we find
$\|\sigma_s\|_{\infty} \le 1$, for all $s>0$.
From the easy bound
$$
\|\sigma_s\|_{\HH_0}\,\le\,\|\sigma_s\|_{\infty}~,
$$
we see that the set $\{\sigma_s~,~s>0\}$ is a bounded subset of $\HH_0$ and
hence weakly sequentially precompact. Therefore there is a 
sequence $s_n\uparrow \infty$ such that
$$ 
\wlim_{n \rightarrow \infty} \sigma_{s_n}=\sigma^{*}
$$
where $\wlim$ denotes the weak limit in $\HH_0$. Since $T^u$ is a
bounded operator for all
$u>0$, we have 
$$ 
\wlim_{n \rightarrow \infty} T^u\sigma_{s_n}=T^u\sigma^{*}~.
$$
We next show
$$
T^u\sigma^{*}\,=\,\sigma^{*}~.
\EQ(tttt)
$$
Indeed,
$$
\eqalign{
T^u \sigma_{s_n}(x)\,&=\,\frac{1}{s_n}\int_{u}^{s_n+u} \dt\, T^t \chi_A(x)\cr
     \,&=\,\sigma_{s_n} - \frac{1}{s_n} \int_{0}^{u}\dt\, T^t \chi_A(x)
          + \frac{1}{s_n} \int_{s_n}^{s_n+u}\dt\, T^t \chi_A(x)~.\cr
}
\EQ(cesaro)
$$
The last two terms in \equ(cesaro) are bounded by $u/s_n$ and
we obtain $T^u \sigma^*=\sigma^*$ for
all $u > 0$ by taking the limit $n \rightarrow \infty$ in
\equ(cesaro). 

Therefore,
$\sigma^*$ is in the eigenspace of the eigenvalue $1$ of $\TTH0$, $t>0$ and so,
by \clm(simplicity)
$\sigma^*=c1$. To compute $c$ we note that $c=(g,\sigma^*)_{\HH_0}$ and,
using the invariance of the measure, we get
$$
\eqalign{
c\,&=\,\lim_{n \rightarrow \infty}
         \left(g,\frac{1}{s_n}\int_{0}^{s_n}\dt\, \TTH0 
         \chi_A\right)_{\HH_0}\cr   
       \,&=\,\lim_{n \rightarrow \infty}\frac{1}{s_n}\int_{0}^{s_n} \dt
             \int \mu(\dx)\, T^t \chi_A = \mu(A)~.\cr
}
$$ 
So we have $c=\mu(A)$ and $\mu(A)=0$ by hypothesis. 
Using this information, we consider
$(\chi_A,\sigma_{s_n})_{\HH_0}$. We have, on one hand,
$$
\lim_{n\rightarrow \infty} (\chi_A,\sigma_{s_n})_{\HH_0}\,=\,
(\chi_A,\sigma^*)_{\HH_0}\,=\,0~,
$$
and on the other hand we have, by \Equ(birerg) and by the dominated
convergence theorem, 
$$
\lim_{n\rightarrow \infty}(\chi_A,\sigma_{s_n})_{\HH_0} \,=\,\int \dx\,
\nu(A)\chi_A Z_0^{-1} e^{-\beta_0 G} > 0~,
$$
and this is a contradiction.
This shows that there is a unique invariant measure for the Markov
process $T^t$ and as a consequence the measure is ergodic.

We will now strengthen the last statement by showing that the invariant
measure is in fact mixing. This will be done by extending the proof
of return to equilibrium given in \CITE[Tr].

\PROPOSITION(mixing) Assume that the conditions of \clm(uniqueness)
are satisfied. Then the invariant measure $\mu(\dx)$ for the Markov
process $T^t$ is mixing, {\it i.e.}, 
$$
\lim_{t\rightarrow \infty} \int \mu(\dx) f(x) T^tg(x)\,=\, 
\int\mu(\dx)f(x) \int\mu(\dx)g(x)~,
$$
for all $f,g \in \L ^2(X,\mu(\dx))$.

\PROOF We denote $\HH = \L ^2(X,\mu(\dx))$ and its scalar product by
$(\cdot,\cdot)\smu$ and by $\|\cdot\|\smu$ its norm.
By \CITE[Yo], Chap XIII.1, Thm 1, $T^t$
defines a contraction semi-group on $\HH$. Since $T^t$ is a strongly
continuous semi-group on $\CC_{\infty}(X)$ (see \CITE[Ne]) and since
$\CC_{\infty}(X)$ 
is dense in $\HH$, we can extend $T^t$ to a strongly continuous
semi-group $\TTHH$ on $\HH$. 
The property of mixing is equivalent to
$$
\wlim_{t\rightarrow \infty} \TTHH f=(1,f)\smu \quad {\rm for~all~~}
f \in \HH~.
\EQ(wlim)
$$
By a simple density argument it is enough to show \equ(wlim) for
a dense subset of $\HH$. 
Let $\CC^2(X)$ denote the bounded continuous function whose first and
second partial derivatives are bounded and continuous. Then---
\CITE[GS], Part II, \para 9---if $f \in \CC^2(X)$, then $\TTHH  f \in \CC^2(X)$
and for any $\tau < \infty$, $\TTHH  f$ is uniformly differentiable w.r.t.
$t\in[0,\tau]$ and 
$$
\frac{\pa}{\pa t}\TTHH  f\,=\, L f~,
$$
where $L$ is the differential operator given in \equ(gen1).
Let $f \in \CC^2(X)$. Using the fact, see \clm(schw) and
\clm(I.posnew), that the 
density of the invariant measure is of the form $h(x)={\widetilde
g}e^{-\beta_0 G/2}$ with $g \in \SS(X)$, we may differentiate under the 
integral and integrate by parts and using the invariance of the
measure we obtain
$$
\frac{\d}{\dt}\|\TTHH f\|\smu^2 \,=\, (L\TTHH f,\TTHH f)\smu+(\TTHH f,L\TTHH f)\smu
\,=\, - 
\kern -1.5em\sum_
{{\scriptstyle i\in \{\L,\R\} \atop 
\scriptstyle m \in\{1,\dots,M\}}
\atop
\scriptstyle \nu \in \{1,\dots,d\}} 
\kern -1em\frac{2\lambda^2_{\im} \gamma_{\im}}{\beta_{i}} 
\| \pa_{r^{(\nu)}_{\im}} \TTHH  f \|\smu^2~,
\EQ(decc)
$$
where $\pa_{r^{(\nu)}_{\im}}$ is the differential operator with domain
$\CC^2(X)$. 
Thus  $\|\TTHH f\|\smu^2$ is decreasing,  bounded below and continuous and so
$\lim_{t\rightarrow \infty}\|\TTHH f\|\smu^2$ exists. As a consequence we find
$$
\| \pa_{r^{(\nu)}_{\im}} \TTHH  f \|\smu^2 \,\in\, \L ^1([0,\infty),dt)~. 
\EQ(l1t)
$$
Following \CITE[Tr] and \CITE[Br], we call a sequence $\{t_n\}$ a
$(*)$-sequence if $t_n
\uparrow \infty $ and
$$\lim_{n \rightarrow \infty}\|\pa_{r^{(\nu)}_{\im}} T_\HH^{t_n} f
\|\smu ^2=0~.
\EQ(starseq)
$$ 
The existence of $(*)$-sequences for our problem 
follows easily from \equ(l1t). 
Further we define an almost $(*)$-sequence as a sequence $s_n
\uparrow \infty $ for which there exists a $(*)$-sequence $\{t_n\}$
with $|s_n-t_n| \rightarrow 0$ as $n \rightarrow \infty$. 
As in \CITE[Tr] we next show that $\wlim_{n \rightarrow \infty} T_\HH^{t_n}f =
\wlim_{n \rightarrow \infty} T_\HH^{s_n}f $. Indeed,
let us choose 
$\tau < \min (t_1,s_1)$. From the inequality
$$
\|\frac{\d}{\dt}\TTHH f\|\smu \,=\,\|T_\HH^{t-\tau} L T_\HH^{\tau}f\|\smu  \,\le\,\|LT_\HH^{\tau}f\|\smu ~,
$$
which holds for $t > \tau$, we have
$$
\|(T_\HH^{s_n}-T_\HH^{t_n})f\|\smu \,\le\, |s_n-t_n|\, \|LT_\HH^{\tau}f\|\smu  \rightarrow 0,
\quad n \rightarrow \infty~,
$$
which shows that $T_\HH^{t_n}f$ and $T_\HH^{s_n}f$ have the same weak limit.

The set $\{\TTHH f, t\ge 0\}$ is bounded and hence sequentially
weakly precompact, so that by passing to a subsequence, we may
assume that
$$
\wlim_{n\rightarrow \infty} T_\HH^{t_n} f\,=\,\gamma~, \quad \gamma \in \HH~.
$$

Next we show that for all $(*)$-sequences $\wlim_{n\rightarrow \infty}
T_\HH^{t_n}$ is a locally constant function $\mu-a.e.$ We begin by showing
that $\gamma$ does not depend on the variables $r$.

Let $\pa^{*}_{r^{(\nu)}_{\im}}$ denote the adjoint of
$\pa_{r^{(\nu)}_{\im}}$
in $\HH$
and let
$\psi \in \CC^{\infty}_{0}(X)$. By the smoothness properties of
the density of the invariant measure we see that the function
$\psi$ is in the domain of 
$\pa^{*}_{r^{(\nu)}_{\im}}$ and we have, using \equ(starseq)
$$
\eqalign{
(\gamma,\pa^{*}_{r^{(\nu)}_{\im}}\psi)\smu \,&=\, \lim_{n\rightarrow \infty}
(T_\HH^{t_n}f, \pa^{*}_{r^{(\nu)}_{\im}}\psi)\smu 
\,=\, \lim_{n\rightarrow \infty}
(\pa_{r^{(\nu)}_{\im}} T_\HH^{t_n}f,\psi)\smu\,=\,0~. \cr
}
$$
Written explicitly,
$$
\int\! \dpp\,\dq\,\dr\, \gamma(p,q,r)
\pa_{r^{(\nu)}_{\im}}\bigl (\psi(p,q,r)h(p,q,r)\bigr )\,=\,0~, 
\EQ(r0)
$$
for any $\psi \in \CC^{\infty}_{0 }(X)$. Since $\gamma \in \HH$,
we may set $\gamma=0$ on the set $A\equiv\{x\in X~;~h(x)=0\}$ and
$\gamma$
is locally integrable and thus defines a distribution in
$\DD'(X)$. By \Equ(r0) the support of the distribution
$\pa_{r^{(\nu)}_{\im}}\gamma(p,q,r)$ does not intersect the set $A$
and thus $\gamma(p,q,r)$ is $\mu$-a.e.\ independent of $r$.

Let $t>0$. Then $\wlim_{n\rightarrow \infty}T_\HH^{t_n+t}f=\TTHH  \gamma$. 
Since $t+t_n \uparrow \infty$, it is easy to show, see \CITE[Br], that
$t_n+t$ has  
an almost $(*)$-subsequence $s_n$ and from the above arguments we
conclude that $\TTHH  \gamma$ is independent of $r$.

Next we show inductively, using Condition ${\bf H2}$ that $\gamma$
does not depend on the variables $p$, $q$.
Let $(\TTHH )^*$ denote the semi-group dual to $\TTHH $ on $\HH$ and denote $Z$ its
generator.

Note that for $\psi \in \CC^{\infty}_{0 }(X)$ we have,
upon integrating by parts
$$
\eqalign{
\frac{\d}{\dt} (\psi,\TTHH f)\smu\,&=\,(\psi,L\TTHH f)\smu \cr
\,&=\,\int\! \dpp\,\dq\,\dr\, \sum_{i\in \{\L,\R\} \atop 
m\in\{1,\dots,M\}} \kern -1em\frac{\lambda^2_{\im} \gamma_{\im}}{\beta_{i}} 
\left[\nabla_{r_\im}\cdot\left(\nabla_{r_\im}+\beta_i
\zi\right){\overline \psi} h \right] \TTHH  f \cr
\,&-\,\int\! \dpp\,\dq\,\dr\,\bigl ( L_0 ({\overline \psi}
h)\bigr )\, \TTHH  f~, 
}
\EQ(rind1)
$$
where $L_0$ is given by
$$
L_0\,=\, \smm r_\Lm \cdot\nabla_{p_1} + \smm r_\Rm \cdot\nabla_{p_n} +
\sum_{j=1}^{n} p_j\cdot\nabla_{q_j} -(\nabla_{q_j}V)\cdot\nabla_{p_j}~.
$$
Since $\CC^{\infty}_{0 }(X)$ is in the domain of $Z$, we get
$$
\displaylines{
\frac{\d}{\dt}(\psi, \TTHH  \gamma)\smu\,=\,(Z \psi, \TTHH  \gamma)\smu 
\,=\,\lim_{n\rightarrow \infty}(Z \psi, LT_\HH^{t_n+t}f)\smu\cr 
\,=\,(Z \psi, \TTHH  \gamma)\smu 
\,=\,-\int\! \dpp\,\dq\,\dr\, \bigl (L_0({\overline \psi}
h)\bigr ) \TTHH  \gamma~. \cr
}
$$
The last equality follows from
\equ(rind1) since $\TTHH  \gamma$ is independent of $r$. 

We next choose $\psi(p,q,r)\in \CC^{\infty}_{0 }(X)$ of the form 
$\psi(r,p,q)=\varphi_1(r)\varphi_2(p,q)h^{-1}(p,q,r)$ with 
$\supp(\varphi_1(r)\varphi_2(p,q)) \cap A=\emptyset$ and $\int\! \dr\, \varphi_1(r)=0$. 
For this choice of $\psi$ we have
$$
(\psi,\TTHH \gamma)\smu\,=\,\int\! \dpp \,\dq\, \bigl (\TTHH \gamma\bigr )(p,q)
\,{\overline\varphi}_2(p,q)\cdot\int\! \dr\, {\overline\varphi}_1(r)\,=\,0~,
$$  
and therefore
$$
\eqalign{
0\,&=\,\int\! \dpp\,\dq\,\dr\,\gamma(p,q)  L_0 ({\overline\varphi}_1(r){\overline\varphi}_2(p,q))
\cr
\,&=\,\int\! \dpp\,\dq\, \gamma(p,q)\nabla_{p_1} {\overline\varphi}_2(p,q)\cdot
\int\!\dr\, \smm r_\Lm {\overline\varphi}_1(r)\cr
\,&+\,\int\! \dpp\,\dq\, \gamma(p,q)\nabla_{p_n} {\overline\varphi}_2(p,q)\cdot
\int\!\dr\, \smm r_\Rm {\overline\varphi}_1(r)~.\cr
}
$$
Since $\varphi_1(r)$ is arbitrary, it follows that
$$
\int\! \dpp\,\dq\, \gamma(p,q)\nabla_{p_1} {\overline\varphi}_2(p,q)\,=\,
\int\! \dpp\,\dq\, \gamma(p,q)\nabla_{p_n} {\overline\varphi}_2(p,q)\,=\,0~,
$$
and thus, by a similar argument as above, $\gamma(p,q)$ must be
$\mu$-a.e.\ independent
of $p_1$ and $p_n$: Thus $\gamma$ is a function $\gamma(p_2,\dots,p_{n-1})$.

Using this information, we choose now 
$$\psi(p,q,r)\,=\,\varphi_1(p_1,p_n)
\varphi_2(p_2,\dots, p_{n-1},q) \varphi_3(r) h^{-1}(p,q,r)~,$$
with 
$ \supp(\varphi_1 \varphi_2 \varphi_3) \cap A=\emptyset$ and 
$\int\!\dpp_1\dpp_n\, \varphi_1(p_1,p_n)=0$. For such a choice of $\psi$
we obtain  
$$
\eqalign{
0\,&=\,\int\! \dpp\,\dq\,\dr\, L_0
({\overline\varphi}_1 {\overline\varphi}_2 {\overline \varphi}_3)
\gamma \cr
\,&=\,\int\! \dpp_2\cdots\dpp_{n-1}\dq\, \gamma 
(\nabla_{q_1} {\overline \varphi}_2) \cdot \int\! \dpp_1\,\dpp_n\, 
p_1\, {\overline \varphi}_1  \int\! \dr\, {\overline \varphi}_3 \cr
\,&+\,\int\! \dpp_2\cdots\dpp_{n-1}\,\dq\, \gamma
(\nabla_{q_n} {\overline \varphi}_2) \cdot \int\!
\dpp_1\dpp_n\,  
p_n {\overline \varphi}_1 \int\! \dr\, {\overline \varphi}_3 ~,\cr
}
$$
and from the arbitrariness of $\varphi_1,\varphi_2,\varphi_3$ we conclude that
$\gamma$ is independent of $q_1,q_n$ (all our statements hold
$\mu$-a.e.).
Finally, choose
$$
\psi(p,q,r)\,=\,\varphi_1(q_1,q_n) \varphi_2(p_2,\dots,p_{n-1},q_2,\ldots,q_{n-1})
\varphi_3(p_1,p_n,r) h^{-1}(p,q,r)~,
$$ 
with $\supp(\varphi_1 \varphi_2 \varphi_3)
\cap A=\emptyset $ and $\int\! \dq_1\dq_n\, \varphi_1(q_1,q_n)=0$. Then we obtain
$$
\eqalign{
0\,&=\,\int\! \dpp_2 \dots \dpp_{n-1} \dq_2 \ldots \dq_{n-1}\, 
\gamma (\nabla_{p_2}{\overline
\varphi}_2)  \cdot \int\!\dq_1\dq_n\,
(\nabla_{q_2}V) 
{\overline \varphi}_1  \int\!\dpp_1\dpp_n\dr\,{\overline
\varphi}_3  \cr 
&+\,\int\! \dpp_2 \dots \dpp_{n-1} \dq_2 \ldots \dq_{n-1}\, 
\gamma (\nabla_{p_{n-1}}{\overline
\varphi}_2) \cdot \int\!\dq_1\dq_n\,
(\nabla_{q_{n-1}}V) 
{\overline \varphi}_1  \int\!\dpp_1\dpp_n\dr\,{\overline
\varphi}_3 ~.\cr  
}
$$
From the arbitrariness of the $\varphi_i$ we conclude in particular that
$$
0\,=\,\int\! \dpp_2 \dots \dpp_{n-1} \dq_2 \ldots \dq_{n-1}\, 
\gamma(\nabla_{p_2}{\overline
\varphi}_2)  \cdot \int\!\dq_1\dq_n\,
(\nabla_{q_2}V ){\overline \varphi}_1~.
\EQ(well)
$$
We may choose $\varphi_1(q_1,q_n)=\pa_{q_1^{(\nu')}}{\widetilde
\varphi}(q_1,q_n)$ for some $\nu' \in \{1,\dots,d\}$ and a positive ${\widetilde
\varphi}(q_1,q_n)$. By Condition {\bf H2} we see 
that 
$$
X^{\nu,\nu'}(q_2)\equiv \int\!\dq_1\dq_n\, (\pa_{q_2^{(\nu)}}V)
 \varphi_1(q_1,q_n)= - \int\!\dq_1\dq_n\,
(\pa_{q_1^{(\nu')}}\pa_{q_2^{(\nu)}}V) {\widetilde \varphi}_1(q_1,q_n)  
$$
is uniformly positive or negative. We can rewrite \equ(well) as
$$
0\,=\,\sum_{ \nu \in \{1,\ldots,d\}} \int\! \dpp_2 \dots \dpp_{n-1}
\dq_2 \ldots \dq_{n-1}\, \gamma \pa_{p_2^{(\nu)}}  X^{\nu,\nu'}(q_2)
{\overline \varphi}_2~,
$$
and we conclude that $\gamma$ is independent of $p_2$. A
similar argument shows that $\gamma$ is independent of $q_{n-1}$
and iterating the above procedure we conclude that $\gamma$
is locally constant $\mu$-a.e. 

So far, we have shown that for all $(*)$-sequences $\{t_n\}$ one has
$\wlim_{n\rightarrow \infty}T_\HH^{t_n}f=\gamma=\const$ From the invariance of
the measure $\mu$ and its ergodicity we conclude that
$$
\gamma=(1,f)\smu\,=\,\int\!\mu(\dx)f(x)~.
$$
We conclude as in [Tr]: suppose that $\wlim_{t\rightarrow
\infty}T_\HH^{t}f\not= (1,f)\smu$. Then by the weak sequential precompactness
of $\{\TTHH  f~;~t\ge 0\}$, there exists a sequence $u_n \uparrow \infty$
for which $\wlim_{t\rightarrow \infty}T_\HH^{t}f= \eta \not= (1,f)\smu$. But,
referring again to \CITE[Br], the sequence $\{u_n\}$ has an almost
$(*)$-subsequence $\{s_n\}$. 
This implies that there is a $(*)$-sequence $\{t_n\}$ such 
that $\wlim_{n\rightarrow \infty}T_\HH^{t_n}f= \eta$. This is a
contradiction, since we have seen that $\wlim_{n\rightarrow
\infty}T_\HH^{t_n}f=(1,f)\smu$ for all $(*)$-sequences. 
By a simple density argument this implies that
$$
\lim_{t\rightarrow \infty} \int \mu(\dx) f(x) \TTHH g(x)\,=\, 
\int\mu(\dx)f(x) \int\mu(\dx)g(x)~,
$$
for all $f,g \in \HH$ and the proof of \clm(mixing) is
complete.

With \clm(mixing) the proof of \clm(2) is now complete.

\SECTION Commutator Estimates and Spectral Properties of $L_{\HH_0}$

In this section, we prove \clm(I.com) and \clm(schw). We generalize the
commutator method of H\"ormander to study the spectral properties of
the operator $L_{\HH_0}$ which is, by 
\clm(I.1), the closure of the differential operator $L$ with domain 
$\CC^{\infty}_{0}(X)$ 
which we defined in
Eq.\equ(gen1). 
We recall the definition:
$$\eqalign{
L\,&=\,\smm{\lambda_\Lm^2 \gamma_\Lm\over \beta_\L}
\left (\nabla_{r_\Lm}-\beta_\L \zl\right ) 
\cdot\nabla_{r_\Lm }\cr 
\,&+\,\smm{\lambda_\Rm^2 \gamma_\Rm\over \beta_\R}
\left (\nabla_{r_\Rm}-\beta_\R \zr\right ) 
\cdot\nabla_{r_\Rm } \cr
  \,&+\, \smm r_\Lm \cdot\nabla_{p_1} + L_\S  +  
\smm r_\Rm \cdot\nabla_{p_n}~, \cr
}
\EQ(gen1repeat)
$$
with the abbreviations
$$
\zl\,=\,\lambda_\Lm^{-2}r_\Lm-q_1~,\quad
\zr\,=\,\lambda_\Rm^{-2}r_\Rm-q_n~,\quad
\EQ(zetadefrep)
$$
and where
$L_\S $ is the Liouville operator associated with the Hamiltonian
$H_\S(q,p)$: 
$$
L_\S \,=\,\sum_{j=1}^{n} p_j\cdot\nabla_{q_j} - 
(\nabla_{q_j}V)\cdot\nabla_{p_j}~.
\EQ(Liouvrep)
$$

For the following estimates it
will be convenient to work in the flat Hilbert space $\L
^2(X,\dx)$. The differential operator $L$ is unitarily equivalent to the operator $\Calkop $ on $\L ^2(X,{\dx})$ with domain $\CC^{\infty}_{0}(X)$ given by
$$
\eqalign{
\Calkop \,&=\,e^{-\beta_0 G/2}\,L\,e^{\beta_0 G/2} \cr
 \,&=\,\alpha -\smm\frac{\lambda_\Lm^2\gamma_\Lm}{\beta_\L}R_\Lm^* R_\Lm
              -\smm\frac{\lambda_\Rm^2\gamma_\Rm}{\beta_\R}R_\Rm^* R_\Rm  
              +\Calkop_\as ~,\cr
}\EQ(positive)$$ 
where $\alpha$ is given by \equ(alpha) and
$$
\eqalign{
R_\Lm\,&=\, \nabla_{r_\Lm } + \sqrt{(\beta_\L -\beta_0/2)\beta_0/2}\,
            \zl ~,\cr
R_\Rm\,&=\, \nabla_{r_\Rm } + \sqrt{(\beta_\R -\beta_0/2)\beta_0/2}\,
           \zr~,\cr
\Calkop_\as \,&=\,  \smm r_\Lm \cdot\nabla_{p_1} + L_\S  + \smm r_\Rm \cdot\nabla_{p_n}~\cr
     &-~~ \frac{\beta_L-\beta_\R}{\beta_\L+\beta _\R }\smm\frac{\lambda_\Lm^2\gamma_\Lm}{2}
         \left(\nabla_{r_\Lm }\cdot\zl 
          +\zl\cdot\nabla_{r_\Lm }
                \right)\cr
      &+~~ \frac{\beta_\L-\beta_\R}{\beta _\L+\beta_\R}\smm\frac{\lambda_\Rm^2\gamma_\Rm}{2}
         \left(\nabla_{r_\Rm }\cdot\zr
            +\zr\cdot\nabla_{r_\Rm }
             \right)~.\cr
}
$$
All subsequent estimates will be valid
for any $f \in \SS(X)$ and thus for all functions in the domain of
$\Calkop$.

It is convenient to introduce the following notations: We introduce
new variables, and recall some earlier definitions: 
Let $n'=[n/2]$ denote the integer part of $n/2$. We define
$$
\eqalign{
P_j\,&=\, \nabla_{p_j}+a_\L p_j ~,\qquad j=1,\dots, n'~,\cr
P_j\,&=\, \nabla_{p_j}+a_\R p_j ~,\qquad j=n'+1,\dots, n~,\cr
Q_j\,&=\, \nabla_{q_j}+a_\L W_j(q,r) ~,\qquad j=1,\dots,n'~, \cr
Q_j\,&=\, \nabla_{q_j}+a_\R W_j(q,r) ~,\qquad j=n'+1,\dots,n~,\cr
R_\Lm\,&=\,\nabla_{r_\Lm } +a_\L W_\Lm(q,r)~,\qquad \mlist~,\cr
R_\Rm\,&=\, \nabla_{r_\Rm }+a_\R W_\Rm(q,r)~,\qquad \mlist~,\cr
}
$$
where
$$
\eqalign{
a_\L\,&=\,\bigl(  (\beta_\L-\beta_0/2)\beta_0/2 \bigr )^{1/2}~,\cr
a_\R\,&=\,\bigl(  (\beta_\R-\beta_0/2)\beta_0/2 \bigr )^{1/2}~,\cr
W_1(q,r)\,&=\,\nabla_{q_1}V(q)-\smm r_\Lm~,\cr
W_j(q,r)\,&=\,\nabla_{q_j}V(q)~,\qquad j=2,\dots,n-1~,\cr
W_n(q,r)\,&=\,\nabla_{q_n}V(q)-\smm r_\Rm~,\cr
W_\Lm(q,r)\,&=\,
           {\lambda_\Lm^{-2}}r_\Lm -q_1~,\qquad \mlist~,\cr
W_\Rm(q,r)\,&=\,
           {\lambda_\Rm^{-2}}r_\Rm -q_n~,\qquad \mlist~.\cr
}
$$

We next define the operators $\Lop_0$, $\Lop $, and $\Lambda $ which
will be used in the statement of our main bound:
$$
\eqalign{
\Lambda \,&=\,\biggl(1+ \sum_{j=1}^n P_j^* P_j +\sum_{j=1}^n Q_j^*
Q_j+\smm \bigl (R_\Lm^*R_\Lm+R_\Rm^*R_\Rm\bigr )\biggr )^{1/2}~,\cr
\Lop_0\,&=\,\Calkop_\as~,\cr
\Lop\,&=\,\alpha -\Calkop\,=\, -\Lop_0
+\smm (b_\Lm R_\Lm^*R_\Lm+ b_\Rm R_\Rm^*R_\Rm)~.\cr
}
\EQ(alldefs)
$$
Here, we use
$$
b_\Lm\,=\,\lambda_\Lm^2\gamma_\Lm/\beta_\L~, \qquad
b_\Rm\,=\,\lambda_\Rm^2\gamma_\Rm/\beta_\R~.
$$

Our main estimate is
\CLAIM Theorem(hor1) Under the Assumptions {\bf H1}, {\bf H2} on $V$,
there are an $\epsilon >0$ and a  $C<  \infty$ such that for all $f\in
\SS(X)$ one has
$$
\| \Lambda^\epsilon  f\|\,\le\, C \lff~.
\EQ(h1)
$$

\PROOF The proof will be an easy consequence
of the following 
\CLAIM Proposition(hor2) There are finite constants
$C_j$, $C_j'$, and $C$ such that for all $f\in \SS(X)$
one has with $n'=[n/2]$,
$$
\eqalignno{
\|\Lambda^{\epsilon_j -1} P_j f\|\,&\le\, C_j \lff~,\quad
j=1,\dots,n'~,\NR(pineq) 
\|\Lambda^{\epsilon_j -1} P_{n+1-j} f\|\,&\le\, C_j \lff~,\quad
j=1,\dots,n-n'~,\NR(pineq2)
\|\Lambda^{\epsilon'_j -1} Q_j f\|\,&\le\, C'_j \lff ~,\quad
j=1,\dots,n'~,\NR(qineq) 
\|\Lambda^{\epsilon'_j -1} Q_{n+j-1} f\|\,&\le\, C'_j \lff ~,\quad
j=1,\dots,n-n'~,\NR(qineq2)
\| R_\Lm f\|+ \|R_\Rm f\|\,&\le\, C\lff~,\quad\mlist~,\NR(rineq)
}
$$
where $\epsilon_j=\quatre^{1-2j}$ and $\epsilon'_j=
\quatre^{-2j}$. 

\LIKEREMARK{Proof of \clm(hor2)} For the $R_\im$, we have the
easy estimate
$$
\|R_\im f\|^2\,=\,(f,R_\im^* R_\im f) \,\le\, b_\im^{-1}\Re (f,\Lop f)
\,\le\, b_\im^{-1}\|\Lop f\|\,\|f\|\,\le\, b_\im^{-1}\lff^2~.
\EQ(startinduct)
$$
This proves Eq.\equ(rineq) for these cases.

For the other cases, the proof will proceed by induction:
It will proceed by bounds on $P_1$, $Q_1$, $P_2,\dots,Q_{n'}$,
and a totally symmetric argument, which is left to the reader, can be
used from the other end of the chain, proceeding over $P_n$, $Q_n$,
$P_{n-1}$,
until the bounds reach the ``center'' of the chain.
We next prepare the inductive proof.
To make the result of this calculation clearer,
we define the matrices
$$
\D_{j,k}\,=\,\nabla_{q_{j}}\nabla_{q_{k}}V~, \quad j,k=1,\dots,n~.
$$
In components, this means, for $\mu,\nu\in \{ 1,\dots ,d\}$,
$$
\D_{j,k}^{(\mu,\nu)}\,=\,\nabla_{q_{j}^{(\mu)}}\nabla_{q_{k}^{(\nu)}}V~, 
\quad j,k=1,\dots,n~.
$$
By our choice of potential $V$ all the $\D_{j,k}$ vanish,
except $\D_{j,j}$, with $j=1,\dots,n$ and $\D_{j+1,j}\,=\D_{j,j+1}$,
with $j=1,\dots,n-1$. Furthermore, by Condition {\bf H1}, all the
$\D_{j,k}^{(\mu,\nu)}$ are uniformly bounded functions of $q$.
Finally, by Assumption {\bf H2}, the matrices $\D_{j,j+1}$ are definite,
with uniformly bounded inverse.

One verifies easily the relations:
$$
\eqalign{
[R_\Lm,\Lop_0]\,&=\,P_1 + c_m R_\Lm~,\cr
[P_1,\Lop  _0]\,&=\,Q_1~,\cr
[Q_1,\Lop_0]\,&=\,-\D_{1,1}P_1-\D_{2,1}P_2+\smm c_m' (R_\Lm+R_\Lm^*)~,\cr
[P_j,\Lop_0]\,&=\,Q_j~,\qquad j=2,\dots,n-1~,\cr
[Q_j,\Lop_0]\,&=\,-\D_{j-1,j}P_{j-1}-\D_{j,j}P_j-\D_{j+1,j}P_{j+1}~,
\qquad j=2,\dots,n-1~,\cr
}\EQ(comm1)
$$
where
$$
\eqalign{
c_m\,&=\,\gamma_\Lm (\beta_\L-\beta_0)/\beta_\L~,\cr
c_m'\,&=\, b_\Lm a_\L  (\beta_\L-\beta_0)~.\cr
}
$$
Symmetrical relations hold at the other end of the chain.
With these notations, we can rewrite (among several possibilities):
$$
\eqalign{
P_1\,&=\,[R_{\L,1},\Lop_0]- c_1 R_{\L,1}~,\cr
Q_1\,&=\,[P_1,\Lop_0]~,\cr
P_2\,&=\,-\D_{2,1}^{-1}\bigl ([Q_1,\Lop_0]+\D_{1,1}P_1-\smm c_m'
(R_\Lm+R_\Lm^*)\bigr 
)~,\cr
Q_j\,&=\,[P_j,\Lop_0]~,\qquad j=2,\dots,n~,\cr
P_{j+1}\,&=\,-\D_{j+1,j}^{-1}\bigl (
[Q_j,\Lop_0]+\D_{j-1,j}P_{j-1}+\D_{j,j}P_j\bigr )~,\qquad j=2,\dots,n'~,\cr
}\EQ(comm2)
$$
with symmetrical relations at the other end of the chain.
We can streamline this representation by 
defining $Q_0=R_{\L,1}$, and
$\D_{1,0}=-1$.
Then we can write, for $j=1,\dots,n'$:
$$
\eqalignno{
P_{j}\,&=\,-\D_{j,j-1}^{-1} \bigl ([Q_{j-1},\Lop_0]+S_j\bigr )~,\NR(inductp)
Q_j\,&=\,\hphantom{-\D_{j+1,j}^{-1} \bigl (}[P_j,\Lop_0]~,\NR(inductq)
}
$$
where the
operators $S_j$
depend linearly on
$\{P_1,\dots,P_{j-1}\}$, $\{Q_1,\dots,Q_{j-1}\}$, and the $R_\Lm$.
The relations Eqs.\equ(inductp) and \equ(inductq)
will be used in the inductive proof.

Such relations are of course
reminiscent of those appearing in the study of hypoelliptic operators.
The novelty
here will be that we obtain bounds which are valid not only in a
compact domain, but in the unbounded domain of the $p$'s and $q$'s.

The following bounds will be used repeatedly:
\CLAIM Proposition(pseudo) Let $Z$ denote one of the operators $Q_j$,
$Q_j^*$,
$P_j$,
or $P_j^*$.
Let $\D$ denote one of the $\D_{j,k}$.
Assume that $\alpha \in(0,2)$. Then the following operators are bounded
in $\L^2(X,\dx)$:
\item{1)}\quad$\Lambda ^{\beta  }[\D,\Lambda ^{-\alpha
}]\Lambda ^\gamma$~, if $\beta +\gamma\le \alpha +1$~,
\item{2)}\quad$\Lambda ^{\beta }Z\Lambda ^{\gamma }$~, if $\beta
+\gamma\le -1$~, 
\item{3)}\quad$\Lambda ^{\beta }[\Lop_0,Z]\Lambda ^{\gamma}$~, if $\beta +\gamma\le -1$~,
\item{4)}\quad$\Lambda ^{\beta } [Z,\Lambda ^{-\alpha }]\Lambda 
^{\gamma }$~,  if $\beta +\gamma\le \alpha +1$~,
\item{5)}\quad$\Lambda ^{\beta }[\Lambda ^{-\alpha },\Lop_0]\Lambda^{\gamma }
$~, if $\beta +\gamma\le \alpha $~.

\PROOF The proof will be given in Appendix B.

Because we are working in an infinite domain, and work with non-linear
couplings,
we will not bound the l.h.s of Eq.\equ(pineq) directly, but
instead the more convenient quantity\footnote{${}^2$}{For readers familiar
with the method of H\"ormander, we 
wish to point 
out that this device seemed necessary because we do not have good
bounds on $[\Lop_0,[Q_1,\Lop_0]]$.}:
$$
R_j(f)\,=\,(\Lambda
^{\epsilon_j-1}\D_{j,j-1}P_jf~,~\Lambda
^{\epsilon_j-1}P_jf )~. 
$$
We have the 
\CLAIM Lemma(q) There is a constant $C$ such that for all
$j\in\{1,\dots,n\}$ and all $f\in \SS(X)$ one has the inequality
$$
\|\Lambda^{\epsilon_{j} -1} P_j f\|^2\,\le\, C\bigl( |R_j(f)|+\|f\|^2 \bigr) ~.
$$

\noindent Therefore, to prove Eq.\equ(pineq), it suffices to prove
the corresponding inequality for the $R_j$.

\LIKEREMARK{Proof of \clm(q)}Let $\D=\D_{j,j-1}$, $\epsilon =\epsilon
_{j}$, and
$P=P_{j}$. Then, by our Assumption {\bf H2}, there is a constant
$m>0$ for which $\D>m$. Therefore,
$$
\eqalign{
\|\Lambda^{\epsilon -1}Pf\|^2\,&=\,(\Lambda^{\epsilon -1}Pf~,~
\Lambda^{\epsilon -1}Pf)\cr
\,&\le\,m^{-1}
|(\D\Lambda^{\epsilon -1}Pf~,~
\Lambda^{\epsilon -1}Pf)|\cr
\,&\le\,m^{-1}|(\Lambda ^{\epsilon -1}\D Pf~,~\Lambda ^{\epsilon
-1}Pf)|
+m^{-1}|([\Lambda ^{\epsilon -1},\D] Pf~,~\Lambda ^{\epsilon
-1}Pf)|\cr
\,&\le\,m^{-1}|R_j(f)|
+m^{-1}\bigl |\bigl ( 
(\Lambda ^\epsilon [\Lambda ^{\epsilon -1},\D]\Lambda )(\Lambda ^{-1}
P)f~,
~(\Lambda ^{-1}P)f\bigr )\bigr |~.\cr
}
$$
The proof of \clm(q) is completed by using the bounds 1) and 2) of
\clm(pseudo). 
\LIKEREMARK{The inductive step}
We begin by the induction step for the $P_j$. 
We assume now that the bounds \equ(pineq)
and \equ(qineq) have been shown for 
all $j\le k$. We want to show \equ(pineq)
for $j=k+1$.
Using Eq.\equ(inductp) and \clm(q), we start by writing
$$
\eqalign{
R_{k+1}(f)\,&\equiv\,
\biggl( \Lambda^{\epsilon_{k+1}-1} \D_{k+1,k} P_{k+1} f~,~ 
 \Lambda^{\epsilon_{k+1}-1} P_{k+1}f\biggr)\cr
\,&=\,
\biggl(\Lambda^{2\epsilon_{k+1} -1}[\Lop_0,Q_k]f~,~ 
\Lambda^{-1}P_{k+1}f\biggr)\cr
&~~\,-\,
\biggl(\Lambda^{2\epsilon_{k+1} -1} S_{k+1}f~,~ \Lambda^{-1}
P_{k+1}f\biggr)\cr
\,&\equiv\, X_1 -X_2~.\cr}
$$
We first bound $X_2$. 
Note that $S_{k+1}$ is a sum of terms of the form $\D T$ where $T$ 
is equal to $P_j$ or $Q_j$ with $j \le k$, and $\D$ is either
a constant or equal to one of the $\D_{k,\ell}$.
Therefore we obtain, 
using \clm(pseudo), the inductive
hypothesis, and the choice $2\epsilon_{k+1}\le \min_{j\le k}(\epsilon
_j,\epsilon '_j)=\epsilon_k'$:
$$
\eqalign{
|(&\Lambda^{2\epsilon_{k+1} -1}
\D T f~,~\Lambda^{-1}P_{k+1}f)|\cr
\,&\le\,
\bigl |\bigl (\D \Lambda^{2\epsilon_{k+1} -1}T f~,~\Lambda
^{-1}P_{k+1}f\bigr )\bigr |
+\bigl |\bigl (([\Lambda^{2\epsilon_{k+1} -1},\D]\Lambda)\,( \Lambda
^{-1}T) f~,~\Lambda^{-1}P_{k+1}f\bigr )\bigr |\cr
\,&\le\,
\OO(1)\lff\|f\| + \OO(1)\|f\|^2
\,\le\,\OO(1) \lff^2~.\cr
}    
$$
This proves the desired bound. 

We now come to the ``interesting'' term $X_1$.
The commutator is rewritten as
$$
\eqalign{
[\Lop_0,Q_k]\,&=\,-Q_k \Lop -\Lop ^*Q_k + 
\HALF\bigl(Q_k (\Lop +\Lop ^*)+(\Lop +\Lop ^*)
Q_k\bigr)\cr
\,&\equiv\,X_3 +X_4 + X_5~.\cr
}
$$
We discuss the 3 corresponding bounds:
\LIKEREMARK{Term $X_3$}In this case, we are led to bound, with 
$\epsilon =\epsilon_{k+1}$, 
$$
\eqalign{
T_3\,&\equiv\,|( Q_k \Lop  f~,~\Lambda^{2\epsilon -2}P_{k+1} f)|\,=\,
|(\Lop  f~,~Q_k^*\Lambda^{2\epsilon-2}P_{k+1} f)|\cr
\,&=\,|(\Lop  f~,
~(Q_k^*\Lambda^{2\epsilon -1})(\Lambda^{ -1}P_{k+1}) f)|\cr
\,&\le\,|(\Lop  f~,
~(\Lambda^{ -1}P_{k+1})(Q_k^*\Lambda^{2\epsilon -1}) f)|+|(\Lop  f~,
~[Q_k^*\Lambda^{2\epsilon -1},\Lambda^{ -1}P_{k+1}] f)| \cr
\,&\equiv\,X_{3,1}+X_{3,2}~.\cr}
\EQ(h5)
$$
We start by bounding $X_{3,1}$. Since $\Lambda^{-1}P_{k+1}$ is
bounded by \clm(pseudo), it suffices to show that 
$$
\|Q_k^* \Lambda^{2\epsilon -1}f\|\,\le\,C\lff~.
\EQ(painful)
$$
To see this we first write, using $Q=Q_k$,
$$
\eqalign{
\|Q^* \Lambda^{2\epsilon -1}f\|^2\,=\,
(f~,~\Lambda^{2\epsilon -1}QQ^*\Lambda^{2\epsilon -1}f)\cr
\,&=\,
\|\Lambda^{2\epsilon -1}Qf\|^2
+
(f~,~[\Lambda^{2\epsilon -1}Q,Q^*\Lambda^{2\epsilon -1}]f)~.
}
$$
The first term is bounded by the inductive hypothesis by $\OO(1)\lff^2$
and the choice
of $\epsilon_{k+1}$, while the second can be bounded by
$\OO(1)\|f\|^2$
by expanding the
commutator (and using \clm(pseudo)):
$$
\displaylines{
[\Lambda^{2\epsilon -1}Q,Q^*\Lambda^{2\epsilon -1}]\,=\,
(\Lambda^{2\epsilon -1}Q^*\Lambda^{-2\epsilon })\Lambda^{2\epsilon}
[Q,\Lambda^{2\epsilon -1}]\cr
+\Lambda^{2\epsilon -1}[Q,Q^*]
\Lambda^{2\epsilon -1}+
\bigl ([
\Lambda^{2\epsilon -1},Q^*]\Lambda^{2\epsilon }\bigr )
\Lambda^{-1}Q~.\cr
}
$$
This proves \Equ(painful).

To bound $X_{3,2}$, we use $[P_{k+1}^*,Q_k]=0$ and we write
$$
\eqalign{
[Q_k^*\Lambda^{2\epsilon -1},\Lambda^{ -1}P_{k+1}]\,&=\,
\bigl(Q_k^*\Lambda^{-1}\bigr)[\Lambda^{2\epsilon -1},P_{k+1}]+
\bigl([Q_k^*,\Lambda^{-1}]\Lambda^{2\epsilon }\bigr)\bigl( \Lambda
^{-2\epsilon }P_{k+1}\Lambda^{2\epsilon -1}\bigr)~.\cr
}
$$
Since each factor above is bounded by \clm(pseudo), the desired bound follows:
$$
T_3\,\le\,\OO(1)\lff^2~.
$$
\LIKEREMARK{Term $X_4$}Here, we want to bound
$
T_4\equiv|( \Lop ^*Q_k f~,~\Lambda^{2\epsilon -2} P_{k+1} f)|
$. We get
$$
\eqalign{
T_4\,=\,|(\Lop ^*Q_k  f~,~\Lambda^{2\epsilon -2}P_{k+1} f)|\,&=\,
|(Q_k f~,~\Lop \Lambda^{2\epsilon-2}P_{k+1} f)|\cr
\,\le\,|(\Lambda^{2\epsilon -1}Q_k f~,~ \Lambda^{-1}P_{k+1}\Lop  f)|
\,&+\,| (Q_kf~,~ [\Lop  ,\Lambda^{2\epsilon -2} P_{k+1}]f)|
\cr\,\equiv\,X_{4,1}&+X_{4,2}~.\cr}
\EQ(h6)
$$
Using the inductive hypothesis, and the bound $\|\Lambda
^{-1}P_{k+1}\|\le\OO(1)$,
the term
$X_{4,1}$ is bounded by 
$$
\|\Lambda^{2\epsilon -1} Q_k f\|\,
\|\Lambda^{-1}P_{k+1}\Lop f\|\,\le\,\OO(1)\lff^2~.
$$
We write the commutator of $X_{4,2}$ as 
$$
\eqalign{
[\Lop ,\Lambda^{2\epsilon -2}P_{k+1}]\,&=\,\Lambda^{2\epsilon
-1}\bigl(\Lambda^{-1}[\Lop ,P_{k+1}]+
\Lambda^{-1} [\Lop_0,\Lambda^{2 - 2\epsilon }]\Lambda^{2\epsilon -1}    
(\Lambda^{-1}P_{k+1})\bigr)~,\cr
}
$$
since $\Lop -\Lop_0$ commutes with $\Lambda $.
Using \clm(pseudo) and the inductive hypothesis this leads to the following
bound for $X_{4,2}$:
$$
\eqalign{
X_{4,2}\,&\le\,|(\Lambda^{2\epsilon -1}Q_kf~,~\Lambda
^{-1}[\Lop ,P_{k+1}]f)|\cr
&~~\,+\,\bigl|\bigl(\Lambda^{2\epsilon -1}Q_kf~,~(\Lambda^{-1} 
[\Lop_0,\Lambda^{2 - 2\epsilon }]\Lambda^{2\epsilon -1})    
(\Lambda^{-1}P_{k+1})f\bigr)\bigr|\cr
\,&\le\,\OO(1)(\|\Lop f\| + \|f\|)(\|f\|)~.
}
$$
This completes the bounds involving $X_4$.
\LIKEREMARK{Term $X_5$}Here, we bound
$$
T_5\,\equiv\, \HALF\biggl(\bigl(Q_k (\Lop +\Lop ^*)+(\Lop +\Lop ^*)
Q_k\bigr)f~,~\Lambda^{2\epsilon -2}P_{k+1}f\biggr)~.
$$
Assume first $k>1$ (and in any case we have $k<n$).
Looking at the definition of $\Lop $, we see that
in this case $Q_k$ commutes with $\HALF(\Lop +\Lop ^*)=\Re \Lop $, and
we can rewrite $T_5$ as
$$
T_5'\,=\,2 \biggl((\Re \Lop )f~,~ Q_k^*\Lambda
^{2\epsilon -2}P_{k+1}f\biggr)
~.
$$
Using the Schwarz inequality and the positivity of $\Re \Lop $, we get a bound
$$
\eqalign{
|T_5'|\,&\le\, \bigl((\Re \Lop ) f~,~ f\bigr)^{1/2}
\bigl((\Re \Lop ) Q_k^* \Lambda^{2\epsilon -2} P_{k+1} f~,~ Q_k^* \Lambda
^{2\epsilon -2} P_{k+1} f\bigr)^{1/2}\cr 
\,&=\,\bigl(\Re( \Lop  f~,~ f)\bigr)^{1/2}
\bigl(\Re( \Lop  Q_k^* \Lambda^{2\epsilon -2} P_{k+1} f~,~ Q_k^* \Lambda
^{2\epsilon -2} P_{k+1} f)\bigr)^{1/2}\cr
\,&=\,\bigl(\Re( \Lop  f~,~ f)\bigr)^{1/2}
\bigl(\Re(\Lambda^{-2\epsilon } \Lop  Q_k^* \Lambda^{2\epsilon -2}
P_{k+1} f~,~ \Lambda^{2\epsilon }Q_k^* \Lambda 
^{2\epsilon -2} P_{k+1} f)\bigr)^{1/2}\cr
\,&\equiv\,\bigl(\Re( \Lop  f~,~ f)\bigr)^{1/2} (\Re (f_1~,f_2))^{1/2}~.\cr
}$$
The first factor is clearly bounded by $\lff^{1/2}$. 
To bound $f_1$, we expand again:
$$
\eqalign{
f_1\,&=\,\Lambda^{-2\epsilon }\Lop  Q^* \Lambda^{2\epsilon -2}Pf\,=\,
(\Lambda^{-2\epsilon }Q^* \Lambda^{2\epsilon -1})(\Lambda^{-1}P) \Lop f\cr
\,&+\,
\Lambda^{-2\epsilon }[\Lop , Q^*] \Lambda^{2\epsilon -2}Pf
\,+\,
\Lambda^{-2\epsilon }Q^*[\Lop , \Lambda^{2\epsilon -2}]Pf\cr
\,&+\,
\Lambda^{-2\epsilon }Q^*\Lambda^{2\epsilon -2}[\Lop ,P]f~.\cr
}
$$
The norm of the first term is bounded by $\OO(1)\lff$.
Using \clm(pseudo), the other terms are bounded by $\OO(1)\|f\|$.
To bound $f_2$ we write
$$
\eqalign{
f_2\,&=\,\Lambda^{2\epsilon }Q^* \Lambda 
^{2\epsilon -2} P f
\,=\,\Lambda^{-1}PQ^* \Lambda^ {4\epsilon -1}f\cr
\,&\,+
\Lambda^{2\epsilon }Q^* \Lambda^{-2\epsilon -1}[\Lambda^{
4\epsilon-1},P]f\cr
\,&+\, \Lambda^{2\epsilon }[Q^*,\Lambda^{-2\epsilon -1}]P\Lambda
^{4\epsilon -1}f~.\cr
}
$$
We control the first term using the inductive hypothesis (it is here
that we use the factor $4\epsilon_{k+1}\le  \epsilon_k'$) and the two
others
by \clm(pseudo).
Combining these bounds, we finally
get the bound $T_5\le\OO(1)\lff^2$, and hence the inequality \equ(pineq)
is shown for all $j$.

It remains to discuss the cases $k=0,1$ for the term $X_5$.
The commutators of $\Re \Lop $ with $Q_0\equiv R_{\L,1}$ or with
$Q_1$ do not vanish and hence
there are additional terms in $T_5'$. They are of the form
$$
\eqalign{
\smm b_\Lm ( [R_\Lm^*R_\Lm, R_{\L,1}]f~,~~\Lambda^{2\epsilon
-2}P_{k+1}f)~,\cr
\smm b_\Lm ( [R_\Lm^*R_\Lm, Q_1]f~,~~\Lambda^{2\epsilon
-2}P_{k+1}f)~.\cr
}
$$
Since $[R_\Lm^*R_\Lm, R_{\L,1}]=\const R_{\L,1}\delta_ {m,1}$ and
$[R_\Lm^*R_\Lm,Q_1]=\const R_\Lm$,
this is obviously bounded by $\OO(1)\lff\,\|f\|$.

We have discussed now all the cases for the inductive bound on the
$P_j$. The discussion of this step for the $Q_j$ is the same, except
that some simplifications appear because of the simpler relations
$Q_j=[P_j,\Lop_0]$. 
The proof of \clm(hor2) is complete.

\LIKEREMARK{Proof of \clm(hor1)}Let $\epsilon \,\le\,\epsilon_\nn'$.
We rewrite
$$
\Lambda^{2\epsilon }\,=\,\Lambda^{2 \epsilon -2}\bigl (1+
\sum_{j=0}^\nn Q_j^*Q_j + \sum_{j=1}^n P_j^* P_j\bigr)~.
\EQ(almostlast)
$$
Note now that for $Q=Q_j$,
$$
\Lambda^{ 2\epsilon -2}Q^*Q\,=\,
 Q^* \Lambda^{ 2\epsilon -2 }Q
+[\Lambda^{2\epsilon  -1},Q^*] Q~.
$$
Using \clm(hor2) and \clm(pseudo), we get a bound
$$
(f~,~\Lambda^{ 2\epsilon -2}Q^*Q f)\,\le\,\OO(1)\lff^2+\OO(1)\|f\|^2~.
$$ 
Of course, the $P$ satisfy analogous relations.
Since $\|\Lambda^{ 2\epsilon -2}f\|\,\le\,\OO(1)\|f\|$, the assertion
\equ(h1) follows by summing the terms in \Equ(almostlast).
The proof of \clm(hor1) is complete.

Using \clm(hor1) we can now prove \clm(I.com). We have
\CLAIM Proposition(compact) If the potential $V$ satisfies Conditions {\bf
H1}, {\bf H2} and if $\beta_0 < 2\min(\beta_\L,\beta_\R)$ there is a
$\lambda^*>0$ such that if the couplings  satisfy $|\lambda_\Lm|$,
$|\lambda_\Rm|\in (0,\lambda^*)$
then both $L_{\HH_0}$ and $L_{\HH_0}^{*}$ have compact resolvent.

\PROOF We show that  the operator $\Calkop$ on $\L^2(X,\dx)$ has
compact resolvent. From \clm(hor1) we get the bound 
$$
\| \Lambda^\epsilon  f\|\,\le\, C \bigl (\| (\Calkop -\alpha -1)
f\|+\|f\|\bigr )~,
\EQ(call)
$$
for all $f \in \SS(X)$. Since, by \clm(I.1),  $\CC^{\infty}_{0}(X)$ is
a core of  $\Calkop$, we see, by taking limits, that the estimate
\equ(call) holds for all $f$ in $D(\Calkop)$.

We note that $\Lambda^2$ has compact resolvent. 
Indeed, recall the definition Eq.\equ(effective) of the effective
Hamiltonian $G$. It is easily checked that, first of all, $G$ grows
quadratically in every direction of $\real^{2d(n+M)}$, for
sufficiently small $|\lambda_\im|$. Second, it is also easily verified
that 
$$
\Lambda ^2\,=\, 1 - \sum_{j=1}^n (\Delta_{p_j} + \Delta_{q_j}) +\smm
(\Delta_{r_{\Lm}} + \Delta_{r_{\Rm}})  +{\bf W}(p,q,r)~,$$ 
and, by construction
$$
\eqalign{
{\bf W}(p,q,r)\,&\approx\,\sum_{j=1}^{n'} a_\L^2 \left ((\nabla_{p_j}G)^2+
(\nabla_{q_j}G)^2\right )
\,+\,\sum_{j=n'+1}^{n} a_\R^2 \left ((\nabla_{p_j}G)^2+
(\nabla_{q_j}G)^2\right )\cr
\,&+\,\smm \left ( a_\L^2 (\nabla_{r_\Lm}G)^2+ a_\R^2 (\nabla_{r_\Rm}G)^2
\right )~,}
$$
up to bounded terms.
Thus ${\bf W}(p,q,r)$ diverges in all directions of $\real^{2d(n+M)}$.
Using the Rellich criterion (see \CITE[RS], Thm XII.67) we conclude that 
$\Lambda ^\epsilon $ has compact resolvent for every $\epsilon >0$.

Therefore, \Equ(call) implies, using again the Rellich criterion, that
$(\Calkop - \alpha -1)^*(\Calkop - \alpha -1)$ has compact resolvent.
We claim this implies that $\Calkop$ itself has compact resolvent.
Indeed, since $\Calkop - \alpha -1$ is strictly $m$-accretive,
its inverse exists, and therefore the operator $\bigl((\Calkop -
\alpha -1)^* (\Calkop - \alpha -1)\bigr )^{-1}=((\Calkop -\alpha
-1)^*)^{-1} (\Calkop - \alpha -1)^{-1}$ exists and is compact. This
implies that $(\Calkop - \alpha - 1)^{-1}$ is compact and hence $\Calkop$
has compact resolvent as asserted.

Finally, we prove \clm(schw). We have the following
\CLAIM Proposition(smooth) Let $g$ denote an eigenvector of $L_{\HH_0}$ or
$L_{\HH_0}^*$. If the assumptions of \clm(I.com) are satisfied 
then $g\exp(\beta_0G/2)$ is in the Schwartz space $\SS(X)$.

\PROOF We prove the corresponding statement for the operator $\Calkop$
on $\L^2(X,\dx)$.
We consider the set of $\CC^{\infty}$ vectors of $e^{\Calkop t}$, i.e the set
$$
\CC^{\infty}(\Calkop)\,\equiv\, \{ f \in \L^2(X,\dx)~;~ 
e^{\Calkop t} \in \CC^{\infty}(\real^+ , \L^2(X,\dx))\}~.
$$
The set $\CC^{\infty}(\Calkop)$ obviously contains all eigenvectors of
$\Calkop$. Therefore \clm(smooth) is a direct consequence of the
following Proposition.

\CLAIM Proposition(cinfty) $\CC^{\infty}(\Calkop)\,=\,\SS(X)$.

\PROOF By Theorem 1.43 in [Da] we have the following characterization of
$\CC^{\infty}(\Calkop)$:  
$$
\CC^{\infty}(\Calkop) \, = \, \cap_{n \ge 0} D(\Calkop^n)~,
$$
where $D(\Calkop^n)=\{ f \in D(\Calkop^{n-1}) , \Calkop^{n-1}f \in
D(\Calkop)\}$. 

Since $\SS(X) \subset D(\Calkop)$ and  $ \Calkop  \SS(X) \subset
\SS(X)$, we have the easy inclusion 
$$
\SS(X)\, \subset\, \cap_n D(\Calkop^n)\,=\, \CC^{\infty}(\Calkop)~.
$$

To show the inclusion in the other direction we will need the
following Theorem which we will prove in 
Appendix C. This is a (slight) generalization of the core Theorem,
[Da], Thm 1.9.

\CLAIM Theorem(core) Let $\BB$ be a Banach space. Let $A:D(A) \rightarrow
\BB$ be m-accretive.  For all $n=1,2,\dots$, if\/ $\DD$ is a subset of
$D(A^n)$ and is
dense in $\BB$ and furthermore $\DD$ is invariant under the semi-group
$e^{A t}$,  
then $\DD$ is a core for $A^n$. 

Given this result we first show that $\SS(X)$ is invariant under
$e^{\Calkop t}$.
For $s\ge0$ we consider the scale of spaces $\NN_{s}$ given
by $\NN_s=D(\Lambda^s)$, with the norm
$\|f\|_{(s)}=\|\Lambda^sf\|$.
For $s\le0$ we let $\NN_s$ be the dual of $\NN_{-s}$.
From the definition of $\Lambda^2$, it is easy to see that 
$\{ \|\cdot\|_{(s)}~;~s=0,1,\cdots \}$ is a system of semi-norms for
the topology of $\SS(X)$ and hence $\SS(X) =\cap_{s}\NN_{s}$.

To show  that $\SS(X)$ is left invariant by the semi-group
$e^{\Calkop t}$ generated by $\Calkop$, it is enough to show that
$$
e^{\Calkop t} \NN_{s} \, \subset \, \NN_{s} \quad {\rm
~for~all~}s \ge 0~.
$$
For $f$, $g$ in $\SS(X)$ we have the identity
$$
\eqalign{
\bigl(\Lambda^{-s}e^{\Calkop^* t}\Lambda^s f~,~ g \bigr)
\,&=\, \bigl(f~,~ \Lambda^{s}e^{\Calkop t}\Lambda^{-s} g \bigr) \cr
\,&=\, \bigl( f~,~ g \bigr) + \int_0^t\!\d\tau 
\bigl(f~,~ \Lambda^{s} \Calkop e^{\Calkop \tau}\Lambda^{-s} g \bigr) \cr
\,&=\, \bigl( f~,~ g \bigr) + \int_0^t\!\d\tau 
\bigl(f~,~ (\Calkop + B)\Lambda^{s} e^{\Calkop \tau}\Lambda^{-s} g \bigr) \cr
\,&=\, \bigl( f~,~ g \bigr) + \int_0^t\!\d\tau 
\bigl(\Lambda^{-s}e^{\Calkop^* \tau}\Lambda^{s}(\Calkop^* + B^*)f~,~g\bigr)~,\cr
}
\EQ(gym)
$$
where
$$
B=[\Lambda^s~,~\Calkop]\Lambda^{-s}~,
$$
is a bounded operator by \clm(pseudo). From \equ(gym) we see
that
$$
\frac{\d}{\dt} \Lambda^{-s}e^{\Calkop^* t}\Lambda^s f\,=\,
\Lambda^{-s}e^{\Calkop^* t}\Lambda^{s}(\Calkop^* + B^*)f ~.
\EQ(fog)
$$
Now $\Calkop^*$ is the
generator of a strongly continuous quasi-bounded semi-group, $B^*$ is
bounded and so, [Ka], Chap.9, Thm 2.7, $\Calkop^* + B^*$ with domain
$D(\Calkop^*)$ is the generator a strongly continuous quasi-bounded
semi-group $e^{(\Calkop^* + B^*)t}$ with $\|e^{(\Calkop^* +
B^*)t}\|\le e^{(\alpha + \|B^*\|)t}$. From \equ(fog) we see that
$$
e^{(\Calkop^* + B^*)t}\,=\,\Lambda^{-s}e^{\Calkop^* t}\Lambda^{s}~.
$$
Thus we obtain
$$
\|\Lambda^{-s}e^{\Calkop^* t}\Lambda^{s}\|\,\le\,e^{(\alpha + \|B^*\|)t}~,
$$
and so $e^{\Calkop^* t }:\NN_{-s}\rightarrow \NN_{-s}$, $s > 0$, is bounded. By
duality $e^{\Calkop t }:\NN_{s}\rightarrow \NN_{s}$, $s > 0$, is also
bounded. This implies that
$$
e^{\Calkop t } \NN_s \, \subset \, \NN_s~, \qquad s>0 ~,
$$
and therefore $\SS(X)$ is invariant under $e^{\Calkop t }$.

We now use \clm(hor1). 
Let $f \in \SS(X)$, then replacing $f$ by $\Lambda^m f$ in \Equ(h1), 
we obtain
$$
\eqalign{
\|f\|_{(m+\epsilon)}\,&\le\, \OO(1) \left(\| \Calkop \Lambda^m f\|
+\|f\|_{(m)}\right)\cr 
                    \,&\le\, \OO(1) \left(\| \Calkop f\|_{(m)} + \|[\Lop
                    ,\Lambda^m]f\| +\|f\|_{(m)} \right)~.\cr
}
$$
Since
$$
\|[\Calkop ,\Lambda^m]f\|\,=\, \|\Lambda^m [\Calkop , \Lambda^{-m}]
\Lambda^m f\| ~,  $$
and since $\Lambda^m [\Calkop ,\Lambda^{-m}]$ is bounded by \clm(pseudo)
we obtain the bound
$$
\|f\|_{(m+\epsilon)} \,\le\, \OO(1) \left(\| \Calkop f\|_{(m)}
+\|f\|_{(m)} \right)~.
\EQ(horm)
$$
Using \equ(horm) it is easy to see, by induction, that, for
$n=1,2,\cdots$ we have
$$
\| f \|_{ (n\epsilon)}\,\le\, \OO(1) \sum_{j=0}^n \left(n \atop j \right) \| \Calkop^j f \|~.
\EQ(nep)
$$
Since $\SS(X)$ is a core for $\Calkop^n$ by \clm(core), we see, by
taking limits, that
$$
D(\Calkop^n)\,\subset\, \NN_{n\epsilon}~.
$$
Therefore
$$
\CC^{\infty}(\Calkop)\,=\,\cap_{n} D(\Calkop^n)\, \subset\, \cap_n  \NN_{n\epsilon}
\,=\, \SS(X)~.
$$
And this concludes the proof of \clm(cinfty).

\def\actualnumber{A}
\SECTIONNONR Appendix A: Proof of \clm(I.1)

If $x(t)=\xi(t,w;x)$ denotes the solution of \equ(sde), it has 
the cocycle property
$$
\xi(t,\tau^s w; \xi(s,w;x))\,=\,\xi(t+s,w; x)~,
$$
which holds for all $t,s \in \real$, $x\in X$ and $w \in \WW $. Here we have
introduced the shift $(\tau^t w)(s)=w(t+s)$ on $\WW$. In particular
the  map $x \mapsto \xi(t,w;x)$ is a bijection with inverse 
$x \mapsto \xi(-t,\tau^t w; x)$. A standard argument shows that these
maps are actually diffeomorphisms (see e.g., [IW], Ch.~V.2).
The Jacobian of 
$\xi(t,w;\cdot)$ is given by
$$
J(t,w;\cdot)\,=\,|{\rm det} D_x \xi(t,w;\cdot)|\,=\,e^{\int_0^t\!\ds\, {\rm
div~}b \circ \xi(s,w;\cdot)} ~,
$$
and according to \equ(div) the Jacobian satisfies
$$
e^{-B|t|}\,\le\, J(t,w;x)\,\le\,e^{B|t|}~.
$$
\REMARK In our case we have in fact
$$
{\rm div~} b\,=\,-d \sum_{i,m} \gamma_{\im}\,\equiv\, -\Gamma < 0~,
$$
so that
$$
J(t,w;\cdot) \,=\, e^{-\Gamma t}~. 
$$
The \clm(I.1) is an immediate consequence of the following lemmata.

\CLAIM Lemma(l2) $T^t$ extends to a strongly continuous, quasi-bounded
semi-group of positivity preserving operators on $\L^2(X,\dx)$. Its
generator is the closure of $L$.
 
\PROOF Let $f\in \CC^{\infty}_{ 0 }$, then we have
$$
\eqalign{
\|T^tf\|^2\,&=\,\lim_{R\rightarrow \infty}\int_{|x|<R}\!\dx\,|T^tf(x)|^2 \,=\,
\lim_{R\rightarrow \infty}\int\!\dx\,\chi_{\{|x|<R\}} \bigl| {\bf E}\bigl[
f\circ \xi(t,w;x) \bigr] \bigr|^2 \cr
\,&\le\,\lim_{R\rightarrow \infty} \int\!\dx\,\chi_{\{|x|<R\}}
{\bf E}\bigl[ |f|^2 \circ \xi(t,w;x) \bigr] \cr
\,&\le\,\lim_{R\rightarrow \infty} {\bf E} \bigl[\int\!\dx\,\chi_{\{|x|<R\}}
  |f|^2 \circ \xi(t,w;x)  \bigr] \cr
\,&\le\,\lim_{R\rightarrow \infty} {\bf E} \bigl[ \int\!\dy\,
\chi_{\{|x|<R\}} \circ \xi(-t,\tau^t w; y) |f|^2 (y) J(-t,\tau^tw;y)\bigr] \cr
\,&\le\,\lim_{R\rightarrow \infty} {\bf E} \bigl[ \int\!\dy\,
1  |f|^2 (y) e^{Bt} \bigr]\,=\,e^{Bt}\|f\|^2 ~.\cr
}
$$
Thus $T^t$ extends to a bounded operator on $\L^2(X,\dx)$ by
continuity. A simple approximation argument shows that $T^t$ is weakly
continuous at $t=0$, and hence, since it is obviously a semi-group,
strongly continuous. Positivity is evident. Now let $f\in
\CC^{\infty}_{0}(X)$, then 
$$
[(1/t)(T^t-1)f-Lf](x)=\int_0^t \ds (1/t)(T^s-1)Lf(x)~,
$$
from which we conclude that the generator
${\widetilde L}$ of $T^t$ on $\L^2(X,\dx)$ satisfies $L\subset
{\widetilde L}$.
 
From the inequality
$
\Re(f,Lf)=-\frac{1}{2} \|\sigma^T \nabla f\|^2-(f~,~{\rm
div~}bf)\le B\|f\|^2,
$
and the fact that $\CC^{\infty}_{ 0 }\subset D(L^*)$ and 
$
\Re(f,L^*f) = -\frac{1}{2} \|\sigma^T \nabla f\|^2+(f~,~{\rm
div~}bf)\le B\|f\|^2,
$
one concludes that:
\item{(i)} $L$ is accretive.
\item{(ii)} The range of $(\lambda-L)$ is dense for $\Re (\lambda)>B$.

Hence, by the Lumer-Phillips Theorem (see for example [Da], Theorem 2.25), the 
closure ${\overline L}$ generates a quasi-bounded semi-group on
$\L^2(X,\dx)$. Since such generators are maximal accretive, we
conclude that ${\widetilde L}={\overline L}$.

We shall now consider the Markov semi-group on weighted $\L^2$-spaces
(such as $\HH_0$)
of the form
$$
\HH_{S}\,=\,\L^2(X,e^{-S}\dx)~,
$$
where $S\in \CC^{\infty}(X)$, and $e^{-S} \in \L^1(X,\dx)$ is
normalized $(\|e^{-S}\|_1 =1)$. We also assume that
$$
b_S\,\equiv\,D\nabla S
$$
with $D$ as in Eq.\equ(thed) satisfies the condition
$$
\| {\rm div~} b_S\|_{\infty}\,<\,\infty~.
\EQ(bdiv)
$$

The action of $T^t$ on $\HH_{S}$ is obviously equivalent to that of 
$e^{-S/2}T^te^{S/2}$ on $L^2(X,\dx)$. For $f\in \CC^{\infty}_{0}(X)$,
Ito's formula gives 
$$
\eqalign{
(e^{-S/2}T^te^{S/2})f(x)\,&=\,{\bf E}\bigl[ e^{(S(x(t))-S(x(0)))/2}
f(x(t))\vert x(0)=x  \bigr] \cr
  \,&=\,{\bf E}\bigl[ e^{\frac{1}{2} \int_0^t\!\ds\,(LS)(x(s)) +
\frac{1}{2} \int_0^t\!\d w(s) (\sigma^T \nabla S)(x(s))}
f(x(t)) \vert x(0)=x  \bigr] \cr
\,&=\,{\bf E}\bigl[ D(t)  e^{\frac{1}{2} \int_0^t\!\ds\, R_S(x(s)) }
f(x(t)) \vert x(0)=x  \bigr]~, \cr
}
$$
where
$$
\eqalign{
D(t)\,&=\,e^{\int_0^t\!\d w(s)\frac{1}{2}(\sigma^T \nabla S)(x(s)) -
\frac{1}{2}\int_0^t\!\ds\,|\frac{1}{2} \sigma^T \nabla S|^2(x(s)) }
\cr 
R_S(x)\,&=\,(LS)(x) + \frac{1}{2}(\nabla S \cdot D \nabla S)(x)\,=\,
{\rm div~}(b_S) + (b+\frac{1}{2}b_S)\cdot \nabla S~. \cr
}
$$
By the Girsanov formula we obtain
$$
(e^{-S/2}T^te^{S/2}f)(x)\,=\,{\bf E} \bigl[e^{\frac{1}{2}
\int_0^t\!\ds\, R_S(y(s))} f(y(t)) \vert y(0)=x \bigr]~,
$$
where $y(t)$ is the Markovian diffusion process defined by the
equation
$$
y(t)\,=\,y(0) + \int_0^t\!\ds\, (b+b_S)(y(s)) + \sigma(w(t)-w(0))~.
$$
Assuming that $R_S$ is bounded above:
$$
\Sigma_S\,=\,\sup_{x\in X} R_S(x) \,<\, \infty~,
$$
and denoting
by $T_S^t$ the semi-group on $\L^2(X,\dx)$ associated with the process
$y(t)$, by \clm(l2) we get
$$
|(e^{-S/2}T^te^{S/2}f)(x)|\,\le\, e^{\frac{1}{2} \Sigma_S t} (T_S^t
|f|)(x)~,
$$
from which one concludes that $e^{-S/2}T^te^{S/2}$ extends to a
strongly continuous, quasi-bounded semi-group of positivity preserving
operators on $L^2(X,\dx)$.

By the Feynman-Kac formula (or Cameron-Martin) we can conclude that the
generator of this semi-group is given, on $\CC^{\infty}_{ 0 }(X)$,
by
$$
L_S\,=\,\nabla\cdot D \nabla + (b+b_S)\cdot\nabla + \frac{1}{2}
R_S\,=\,e^{-S/2}Le^{S/2}~.
$$
Repeating the argument of \clm(l2) we conclude that ${\overline L_S}$
is the generator. Since $\CC^{\infty}_{ 0 }(X)$ is invariant by
$e^{\pm S/2}$ we get

\CLAIM Lemma(l2w) Let $S\in \CC^{\infty}(X)$ be such that
\item{(i)}$b_S=D\nabla S$ satisfies Condition \equ(bdiv),
\item{(ii)}$\sup_{x\in X} (b+\frac{1}{2}b_S)\cdot \nabla S (x) <
\infty$.\hfill\break
Then the semi-group $T^t$ extends to a strongly continuous
quasi-bounded semi-group on $\HH_S$. Moreover $\CC^{\infty}_{0}(X)$ is
a core for its generator.

Now \clm(I.1) is a direct consequence of \clm(l2w). Indeed we apply
\clm(l2w) to
$$
S(x)\,=\,\beta_0 G(p,q,r)~,
$$
where $G$ is given by \equ(effective). 
We see that Condition (i) of \clm(l2w) is satisfied, since $G(p,q,r)$
is of the form quadratic + bounded. An explicit computation shows
that the assumption
$$
\beta_0 \, < \, 2 \min (\beta_\L , \beta_\R)~,
$$
implies that Condition (ii) of \clm(l2w) is satisfied and that the
semi-group satisfies the bound $\|T^t\|\le e^{\alpha t}$ where $\alpha$ is
given by \Equ(alpha).

\def\actualnumber{B}
\SECTIONNONR Appendix B: Proof of \clm(pseudo)

To prove the claims it is useful to introduce some machinery
which replaces the pseudo-differential calculus, which seems
unavailable for the class of operators we want to consider. This may
be useful in its own right.

Let $\FF$, as in the Hypotheses {\bf H1}, {\bf H2}
denote the class of functions of $q\in\real^{dn}$ which are
bounded together with all their derivatives. Let $\YY$ denote the
linear space of operators spanned by
$$
f(q)q^m\partial_q^{m'} p^n\partial_p^{n'}r^\ell\partial_r^{\ell'}~,
\EQ(list)
$$
where $f\in\FF$. (The notation is sloppy, we really
mean components $\nu=1,\dots,d$ of each of these quantities.) We shall
say that the quantities in Eq.\equ(list) are of degree $m+m'+n+n'+\ell+\ell'$.
We let $\YY^s$ denote the subspace of $\YY$
spanned by the expressions of degree less than or equal to $s$.
Clearly, the operators $Z$ and $\MM$ of \clm(pseudo) are in $\YY^1$ and
$\YY^0$, respectively, while $\Lop_0$ and $\Lambda ^2$ are in $\YY^2$.
Below, we shall use this, but also
an additional property of
the potential $V$.
We have
\CLAIM Lemma(lamclam) One has the inclusion
$$
[\Lambda ^2,\YY^s]\,\subseteq\, \YY^{s+1}~.
\EQ(lamclam)
$$
Furthermore, $Y\in\YY^0$ defines a bounded operator.

\PROOF By inspection.

\CLAIM Proposition(bs) Assume that $Y\in\YY^j$, for some
$j\in\{0,1,\dots\}$.
Then
$$
\Lambda ^\beta Y\Lambda ^\gamma
\EQ(bs1)
$$
defines a bounded operator
on $\L ^2(X,\dx) $, when
$$
\beta +\gamma\,\le\, -j~.
$$
Let $Z$ be an operator in $\YY$.
Assume that $[\Lambda ^2,Z]\in \YY^j$, for some
$j\in\{0,1,\dots\}$. Then
$$
\Lambda ^\beta  [\Lambda ^{-\alpha },Z]\Lambda ^\gamma
\EQ(bs2)
$$
defines a bounded operator on $\L ^2(X,\dx) $
for all $\alpha $, $\beta $ and $\gamma$
satisfying
$$
\beta +\gamma\,\le\,\alpha -j +2~.
$$

We will give bounds on various quantities involving $\YY^s$. For this,
we will use throughout the following device:
\CLAIM Lemma(intbound) Let $A_z$ be a bounded continuous
operator-valued function of $z$ and let $F(\lambda,z)$ be a real,
positive continuous bounded function. Then 
$$
\| \iii A_z F(\Lambda ,z) u \| \,\le\,
\sup_{y\ge 0} \|A_y\|\, \|u\| \iii \sup_{\lambda\ge1}F(\lambda,z)~.
\EQ(b3)
$$
If furthermore $A=A_z$
is independent of $z$, one has the bound
$$
\| \iii A F(\Lambda ,z) u \|\,\le\,
\|A\|\, \|u\|  \sup_{\lambda\ge1}\iii F(\lambda ,z)~.
\EQ(b4)
$$

\PROOF Note first that 
$$
\eqalign{
\| \iii A_z F(\Lambda ,z) u \|
\,\le\,
\iii\|  A_z\|\,\| F(\Lambda ,z) u \|~.
}
$$
Since $\Lambda $, as an
operator, satisfies $\Lambda \ge 1$ we also have
from the spectral theorem:
$$
\| F(\Lambda ,z) \|\,\le\, \sup_{\lambda\ge1}F(\lambda,z)~.
$$
Thus, Eq.\equ(b3) follows. In a similar way:
$$
\| \iii A F(\Lambda ,z) u \| \,\le\,
\|A\|\, \|\iii F(\Lambda ,z)\|\,\|u\|\,\le\,
\|A\|\, \|u\|  \sup_{\lambda\ge1}\iii F(\lambda ,z)~,
$$
which is \equ(b4). The proof of \clm(intbound) is complete.

We shall also make use of the following identity, 
valid for $\alpha \in(0,2)$, [Ka] Thm.~V.3:
$$
\Lambda^{-\alpha }\,=\,{\sin (\pi \alpha/2) \over \pi}
\iii z^{-\alpha/2 }(z+\Lambda ^2)^{-1}~.
\EQ(insint)
$$
We also let $C_{-\alpha} ={\sin (\pi \alpha/2) /\pi}$.
\LIKEREMARK{Proof of \clm(bs)}It is obvious that if we show the claim
for $\beta +\gamma=-j$, then it also follows for $\beta +\gamma<-j$.
By the definition of $\YY^j$, and
observing that $f(q)$ is 
bounded, and
by the explicit form of
$\Lambda^{2}$,
we see that the claim holds when $\gamma=0$.
We next consider the case $\beta +\gamma\le0$, $-1\le \gamma<0$. In this case
we write
$$
\Lambda ^{\beta} Y \Lambda ^{\gamma}\,=\, Y \Lambda ^{\beta+\gamma} + \Lambda ^{\beta}
[Y,\Lambda ^{\gamma}]~.
$$
The first term is clearly bounded as in the case $\gamma=0$, by considering
adjoints.
The second term can be written as
$$
\eqalign{
\Lambda^{\beta}[Y,\Lambda^{\gamma}]\,&=\,C_{\gamma}\iii z^{\gamma/2} \Lambda^{\beta} \left [ Y, {1\over
z+\Lambda^{2}}\right ]\cr
\,&=\,C_\gamma\iii z^{\gamma/2} {\Lambda^{\beta}\over z+\Lambda^{2}} [Y,\Lambda^{2}]
{1\over z+\Lambda^{2}}~.\cr
}
\EQ(ex1)
$$
By \clm(lamclam), we see that $[Y,\Lambda^{2}]\in \YY^{j+1}$ and thus,
we get, using Eq.\equ(b3),
$$
\bigl \|\Lambda^{\beta}[Y,\Lambda^{\gamma}]\bigr \|
\,\le\,C_\gamma\sup_{y\ge0}  \bigl\|{\Lambda^{\beta}\over y+\Lambda^{2}}
[Y,\Lambda^{2}]\bigr \|\,\,
\iii z^{\gamma/2}\sup _{\lambda \ge1} {1\over z+\lambda^2}~.
$$
The norm is bounded because $\gamma\in[-1,0)$ and thus
$\beta -2+j+1\le \beta +\gamma+j=0$. The sup over $\lambda$ is
$(1+z)^{-1}$ and the integral converges because $\gamma\in[-1,0)$.

We now proceed to the other choices of $\gamma$ by induction. We first deal
with negative $\gamma$. Assume we have shown that
$\Lambda ^\beta Y\Lambda ^{\gamma'}$ is bounded for all
$\gamma'\in[-\tau ,0]$,
and assume that $\gamma\in[-\tau-1,-\tau)$, and $Y\in\YY^{j}$. We write
$$
\Lambda^{\beta}Y\Lambda^{\gamma}\,=\,\Lambda^{\beta-1}Y\Lambda^{\gamma+1}+
\Lambda^{\beta}[Y,\Lambda^{-1}]\Lambda^{\gamma+1}~.
$$
The first term is bounded by the inductive hypothesis. To bound the
second, we apply again the method used in Eq.\equ(ex1). Then we get
$$
\eqalign{
\Lambda^{\beta}[Y,\Lambda^{-1}]\Lambda^{\gamma+1}
\,&=\,
C_{-1}\iii z^{-1/2} {\Lambda^{\beta}\over z+\Lambda^{2}}
[Y,\Lambda^{2}] {\Lambda^{\gamma+1}\over z+\Lambda^{2}}~.\cr
}
$$
Since $[Y,\Lambda^{2}]\in\YY^{j+1}$, we see from the inductive
hypothesis
that 
$$
\sup_{y\ge0}\left\|\Lambda ^\beta {1\over y+\Lambda ^2} [Y,\Lambda ^2]\Lambda
^{\gamma+1}\right \|\,\le\,\left\|\Lambda ^{\beta-2} [Y,\Lambda ^2]\Lambda
^{\gamma+1}\right \|
$$
is bounded and hence, using \equ(b3), we can complete  the
inductive step.

The case $\gamma>0$ is handled by observing that $Y\in \YY^{j}$
implies
$Y^*\in\YY^{j}$, and bounding
$\Lambda^{\beta}Y\Lambda^{\gamma}$
by bounding $\Lambda^{\gamma}Y^*\Lambda^{\beta}$.
This completes the proof of the first part of \clm(bs).
 
To prove the second part, we first consider the case
$\alpha \in (0,2)$.  
In this case, using Eq.\equ(insint),
we write $\Lambda ^\beta [\Lambda^{-\alpha },Z]\Lambda ^\gamma$ as
$$
\eqalign{
&C_{-\alpha}\int_0^\infty \!\d z\,z^{-\alpha/2}
\Lambda^\beta \left [
{1\over z+\Lambda ^2}~,~Z
\right ]\Lambda^\gamma
\,=\,C_{-\alpha}\int_0^\infty \!\d z\,z^{-\alpha/2}
{\Lambda^\beta  \over z+\Lambda ^2}\left [
{\Lambda ^2}~,~Z
\right ]{\Lambda^\gamma \over z+\Lambda ^2}~.\cr
}
\EQ(calpha)
$$
We let $B=[\Lambda ^2,Z]$, use another commutator and rewrite \equ(calpha) as
$$
\eqalign{
&\,C_{-\alpha}\int_0^\infty \!\d z\,z^{-\alpha/2}
\Lambda^\beta  B {\Lambda ^{\gamma }\over (z+\Lambda ^2)^2}
\,+\,
C_{-\alpha}\int_0^\infty \!\d z\,z^{-\alpha/2}
{\Lambda ^\beta \over z+\Lambda ^2} \bigl [
\Lambda ^2, B
\bigr ]
 {\Lambda ^{\gamma }\over (z+\Lambda ^2)^2}\cr
\,&\equiv\, C_{-\alpha}(X_1 + X_2)~.
}
$$
We first bound $X_1$.
We get, using Eq.\equ(b4),
$$
\eqalign{
\|X_1 u\|\,&=\,
\left \|\Lambda^\beta B\Lambda^{-j-\beta } \iii z^{-\alpha /2
}{\Lambda ^{j+\beta +\gamma }\over 
(z+\Lambda ^2)^2}u\right \|\cr
\,&\le\,\|u\|\,\left \|\Lambda^\beta B\Lambda^{-j-\beta }\right \|\,
\,\sup_{\lambda\ge1}\iii z^{-\alpha /2 
}{\lambda ^{j+\beta +\gamma }\over 
(z+\lambda ^2)^2}\cr
\,&\le\,\|u\|\,\left \|\Lambda^\beta B\Lambda^{-j-\beta }\right \|\,
\,\sup_{\lambda\ge1}\int_0^\infty \d s \, {s}^{-\alpha /2 
}\lambda^{2-\alpha }{\lambda ^{j+\beta +\gamma }\over 
(s+1)^2\lambda^4}~.\cr
}
$$
Since, by assumption, $B\in\YY^j$, the norm is bounded, and the integral is bounded
because
$\beta +\gamma\le\alpha -j+2$, by assumption.

To bound $X_2$ we first observe that by assumption, and by
\clm(lamclam),
$C=[\Lambda ^2,B]$ is in $\YY^{j+1}$.
Therefore, using Eq.\equ(b3) we find the following bound for $X_2$:
$$
\eqalign{
\|X_2u\|\,&=\,\left \|\iii z^{-\alpha/2}
{\Lambda ^\beta \over z+\Lambda ^2} 
C\Lambda ^{1-j-\beta }
\cdot {\Lambda ^{j-1+\beta +\gamma }\over (z+\Lambda ^2)^2}u\right \|\cr
\,&\le \,\|u\| \sup_{y\ge0}\left \|{\Lambda ^\beta \over y+\Lambda ^2} 
C\Lambda ^{1-j-\beta }\right \|\,\,
 \iii z^{-\alpha/2}\sup_{\lambda\ge1}
 {\lambda ^{j-1+\beta +\gamma }\over (z+\lambda ^2)^2}~.\cr
}
$$
This is clearly bounded when $\beta +\gamma\le\alpha -j+2$ and $\alpha
\in(0,2)$. 
This completes the second part of \clm(bs) when $\alpha \in(0,2)$. If
$\alpha =0$ the assertion is trivial. 
The case $\alpha =2$ is handled by considering the identity:
$$
[\Lambda^{-2},Z]\,=\,\Lambda ^{-2} [\Lambda ^2,Z]\Lambda ^{-2}~.
$$
The cases when $\alpha >2$
follow inductively by using the identity:
$$
\Lambda ^\beta [\Lambda ^{-\alpha-2 },Z]\Lambda ^\gamma\,=\,
\Lambda ^{\beta-2} [\Lambda ^{-\alpha },Z]\Lambda ^\gamma+
\Lambda ^{\beta } [\Lambda ^{-2 },Z]\Lambda ^{\gamma-\alpha}~.
$$
The cases when $\alpha <0$ follow by similar identities
The proof of \clm(bs) is complete.

\LIKEREMARK{Proof of \clm(pseudo)}The \clm(pseudo) is a simple
consequence of \clm(bs). Since $Z$ is in $\YY^1$, the claim 2)
is covered by the bound on \equ(bs1). We next prove 3). The operator
$\Lop_0$ is in 
$\YY^2$ and $Z$ is in $\YY^1$. Power counting would suggest that
$[\Lop_0,Z]\in\YY^{2}$. However, by Condition {\bf H1}, we know that
$\nabla
_{q_j}V$ equals a
linear term plus a term in $\YY^0$, and hence explicit computation
shows that $[\Lop_0,Z]\in\YY^1$. Hence the assertion is covered by the
bound on \equ(bs1). Since $\MM\in\YY^0$ we see from \clm(lamclam) that
$
[\MM,\Lambda ^2]\in\YY^1$, and therefore the claim 1) follows from the
bound on \equ(bs2). Using again explicit calculation and Condition
{\bf H1} we see that $[Z,\Lambda ^2]$ is in $\YY^1$ (and not only in
$\YY^2$) and
$[\Lop_0,\Lambda ^2]$ is in $\YY^2$ (and not only in $\YY^3$), and
hence the cases 4) and 5) follow by applying again the bound on \equ(bs2).
The proof of \clm(pseudo) is complete.
\def\actualnumber{C}
\SECTIONNONR Appendix C: A Generalized Core Theorem
 
We prove here the following result from Section 4:
\LIKEREMARK{\clm(core)}{\sl Let $\BB$ be a Banach space. Let $A:D(A) \rightarrow
\BB$ be m-accretive.  For all $n=1,2,\dots$, if\/ $\DD$ is a subset of
$D(A^n)$ and is
dense in $\BB$ and furthermore $\DD$ is invariant under the semi-group
$e^{A t}$,  
then $\DD$ is a core for $A^n$.} 

\PROOF Let $\norm  f \norm _n = \sum_{j=0}^{n}\|A^j f \|$. Then one has
\item{(i)} $(D(A^n) , \norm  \cdot \norm _n)$ is complete.
\item{(ii)} $e^{A t}$ is a strongly continuous semi-group on $(D(A^n),
\norm  \cdot \norm _n)$.

The statement of the theorem is equivalent to the following: If ${
\overline \DD}^n$ denotes the closure of $\DD$ in the norm $\norm  \cdot
\norm _n$, then we have
$$
{\overline \DD}^n\,=\,D(A^n)~.
\EQ(clon)
$$
We show this by induction. For $n=1$, this is the core
Theorem, [Da], Thm 1.9. Let us assume that \equ(clon) holds for
$n-1$. Let $f\in D(A^n)$, so there is a sequence $\{f_m\} \in \DD$
such that 
$$
\lim_{m \rightarrow \infty} \norm  f_m - f \norm _{n-1}\,=\,0~. 
$$
With the $f_m$ we construct a sequence which converges to $f$ in the 
$\norm  \cdot  \norm _{n}$ norm.
We set 
$$
f^{(n,t)}_m \,=\,\int_0^t\!\ds\, \frac{(t-s)^{n-1}}{(n-1)!}\, e^{As} f_m ~.
$$ 
By the above property (ii), $e^{A s}$ is strongly continuous in $s$ in
$D(A^n)$ and hence $f^{(n,t)}_m \in { \overline \DD}^n$.
We set 
$$
f^{(n,t)} \,=\,\int_0^t\!\ds\, \frac{(t-s)^{n-1}}{(n-1)!}\, e^{As} f ~.
$$
Since $e^{A s} g$ is $n$-times differentiable in $s$ when $g \in D(A^n)$, we
obtain, upon integrating by parts:
$$
\eqalign{
\norm  f^{(n,t)}_m - f^{(n,t)}  \norm _n 
\, &= \, \sum_{j=0}^{n-1} \biggl\| \int_0^t\!\ds\, 
\frac{(t-s)^{n-1}}{(n-1)!}\, e^{As} A^j (f_m - f) \biggr\| \cr
\, &+ \, \biggl\| \int_0^t\!\ds\, \frac{(t-s)^{n-1}}{(n-1)!}\, e^{As} A^n
(f_m - f) \biggr\| \cr 
\, &= \, \sum_{j=0}^{n-1} \biggl\| \int_0^t\!\ds\, 
\frac{(t-s)^{n-1}}{(n-1)!}\, e^{As} A^j (f_m - f) \biggr\| \cr
\, &+ \, \biggl\| - \frac{t^{n-1}}{(n-1)!}\, A^{n-1} (f_m - f) +
\int_0^t\!\ds\,  \frac{(t-s)^{n-2}}{(n-2)!}\, e^{As} A^{n-1} (f_m - f)
\biggr\| \cr 
\,&\le\, C e^{\gamma t} \frac{t^{n}}{n!} \norm 
 f_m-f \norm _{n-1} + (C e^{\gamma t} + 1) \frac{t^{(n-1)}}{(n-1)!} \|
A^{n-1}(f_m-f) \| \cr
\, &= \, o(1) \quad {\rm ~as~} m \rightarrow \infty~,
}
$$
by the inductive hypothesis. This shows that $f^{(n,t)} \in {
\overline \DD}^n$.

To conclude we show that $\norm  \frac{n!}{t^n} f^{(n,t)} - f \norm_n 
\rightarrow 0$ as $t\rightarrow 0$. We have
$$
\eqalign{
\bnorm  \frac{n!}{t^n} f^{(n,t)} - f \bnorm _n
\,&=\, \sum_{j=0}^{n-1} \biggl\| \frac{n!}{t^n} \int_0^t\!\ds\,
\frac{(t-s)^{n-1}}{(n-1)!} (e^{As}-1) A^j f \biggr\| \cr
\, &+ \,
\biggl \| \frac{n!}{t^n} \bigl( \int_0^t\!\ds\, \frac{(t-s)^{n-1}}{(n-1)!}
e^{As} A^n f - \frac{t^n}{n!} A^n f \bigr) \biggr\|~.
}
$$
Using that $e^{At} f$ is $n$-times differentiable in $t$, that 
$$
\int_0^t\!\ds\, \frac{(t-s)^{n-1}}{(n-1)!} e^{As} A^n f~,
$$
is the remainder term in the Taylor expansion of $e^{At} f$, and that
for $g\in D(A)$
$$
(e^{As}-1)g = \int_0^s\!\d u\, e^{Au} A g~,
$$ 
if $g\in D(A)$ we obtain
the bound
$$
\eqalign{
\bnorm  \frac{n!}{t^n} f^{(n,t)} - f \bnorm _n 
\, &= \,
\sum_{j=0}^{n-1} \biggl\| \frac{n!}{t^n}
\int_0^t\!\ds\,\frac{(t-s)^{n-1}}{(n-1)!}  \int_0^s\!\d u\, e^{Au}
A^{j+1} f \biggr\| \cr 
\, &+ \, 
\biggl\| \frac{n!}{t^n} \bigl ( e^{At}f - f - Af - \cdots - \frac{t^n}{n!}
A^n f \bigr) \biggr\| \cr
\,&\le\, \OO(t) + o(1)~.
}
$$
This shows that $f \in { \overline \DD}^n$. This proves that 
${\overline \DD}^n = D(A^n)$ as required.

\LIKEREMARK{Acknowledgments}We thank P. Collet for helpful remarks
concerning \clm(uniqueness). We have also profited from useful
discussions with G. Ben Arous, G. Gallavotti, B. Helffer, V. Jak\v
si\'c,
H.M. Maire,
Ch. Mazza, H. Spohn, and  A.-S. Sznitman.
This work was partially supported by the
Fonds National Suisse. 

\LIKEREMARK{References}

\widestlabel{[XXX]}
{\eightpoint

\ITEM[Br]
  \REF       {Bruck, R.E.}
             {Asymptotic Convergence of Nonlinear Contraction
             Semigroups on Hilbert Space}
             {J. Funct. Anal.} 18 15 1975

\ITEM[CL]
  \REF       {Casher,~A. and~J.L.~Lebowitz}
             {Heat Flow in Regular and Disordered Harmonic Chains}
             {\JMP} 12 1701 1971
  
\ITEM[Da]
 \BOOK       {Davies,~E.B.}
             {One-Parameter Semigroups}
             {London}
             {Academic Press}
             {1980}


\ITEM[FK]
  \REF        {Ford,~G.W. and ~M.~Kac}
              {On the Quantum Langevin Equation}
              {\JSP} 46 803 1987

\ITEM[FKM]
  \REF        {Ford,~G.W., M.~Kac and P.~Mazur}
              {Statistical Mechanics of Assemblies of Coupled Oscillators}
              {\JMP} 6 504 1965

\ITEM[GC]
 \REF         {Gallavotti,~G. and E.G.D.~Cohen}
              {Dynamical Ensembles in Stationary States}
	      {\JSP} 80 931 1995

\ITEM[GS] 
   \BOOK      {Gihman,~I.I. and~A.V.~Skorohod}
	      {Stochastic Differential Equations} 
              {Berlin}   
              {Springer} 
              {1972}

\ITEM[He]     
  \BOOK       {Helffer~B.}
              {Semi-Classical Analysis for the Schr\"odinger Operator
               and Applications}
              {Lecture Notes in Mathematics {\bf 1336},  Berlin}   
              {Springer} 
              {1988}

\ITEM[HM]     
  \REF        {Helffer~B. and A.~Mohammed}
              {Caract\'erisation du Spectre Essentiel de l'Op\'erateur
               de Schr\"odinger avec un Champ Magn\'etique}
              {Ann. Inst. Fourier} 38 95 1988

\ITEM[HN]     
  \BOOK       {Helffer~B. and J.~Nourrigat}
              {Hypoellipticit\'e Maximale pour des Op\'erateurs
               Polyn\^omes de Champs de Vecteurs}
              {Boston}
              {Birkh\"auser} 
              1985

\ITEM[H\"o] 
   \BOOK      {H\"ormander~L.}
	      {The Analysis of Linear Partial Differential Operators I-IV}
              {Berlin}   
              {Springer}
              {1985} 
	
\ITEM[Ho]
   \BOOK      {Hopf~E.}
              {Ergodentheorie}
              {Berlin}
              {Springer}
              {1970  reprint}
       
\ITEM[IW]
  \BOOK       {Ikeda~N. and S~.Watanabe}
              {Stochastic Differential Equations and Diffusion
              Processes}
              {Amsterdam}
              {North-Hol\-land}
              {1981}

\ITEM[JP1]
  \REF        {\JP}
              {Ergodic Properties of the non-Markovian Langevin Equation}
              {\LMP} 41 49 1997

\ITEM[JP2]
  \TOAPPEAR   {\JP}
              {Ergodic Properties of Classical Dissipative Systems I}
              {Acta Mathematica}

\ITEM[JP3]
  \TODO       {\JP}
              {Ergodic Properties of Classical Dissipative Systems II}
             
\ITEM[JP4]
  \REF  {\JP}
              {Spectral Theory of Thermal Relaxation}
              {\JMP} 38 1757 1997 

\ITEM[Ka] 
  \BOOK       {Kato,~T.}
	      {Perturbation Theory for Linear Operators}
              {Berlin}
              {Springer} 
              {1980}

%

\ITEM[Ne] 
   \BOOK      {Nelson,~E.}
	      {Dynamical theories of Brownian Motion}
              {Princeton}
              {Princeton university press} 
              1980

\ITEM[OL]
  \REF        {O'Connor,~A.J. and~Lebowitz,~J.L.}
              {Heat Conduction and Sound Transmission in Isotopically 
               Disordered Harmonic Crystals}
              {\JMP} 15 692 1974

%

\ITEM[R-B]     
  \TODO       {Rey-Bellet~L.}
              {Steady States and Transport Properties for Mechanical
              Systems Coupled to Stochastic Heat Baths}

\ITEM[RLL]  
  \REF        {Rieder,~Z.,~Lebowitz,~J.L. and~E.Lieb}
              {Properties of a Harmonic Crystal in a Stationary Nonequilibrium
               State}
              {\JMP} 8 1073 1967

\ITEM[RS] 
   \BOOK      {Reed~M. and~B.~Simon}
	      {Methods of Modern Mathematical Physics I-IV}
              {Boston}   
              {Academic Press} 
              1978


\ITEM[SL]   
   \REF       {Spohn,~H. and~J.L.~Lebowitz}
	      {Stationary Non-Equilibrium States of Infinite Harmonic Systems}
              {\CMP} 54 97 1977  
              
\ITEM[Tr]
  \REF        Tropper,~M.M.
              {Ergodic and Quasi-deterministic Properties of
              Finite Dimensional Stochastic Systems}
              {\JSP} 17 491 1977

\ITEM[Yo] 
   \BOOK      {Yosida,~K.}
	      {Functional Analysis}
              {Berlin}   
              {Springer} 
              {1980}

}

\bye